%% file: main.tex
\documentclass[a4paper,11pt]{article}
\pdfoutput=1 % if your are submitting a pdflatex (i.e. if you have images in pdf, png or jpg format)
\input{packages}
\input{header}
\begin{document}
\maketitle
\flushbottom

\input{intro}
\input{design}
\input{detectorresponse}

\input{setup}
\input{summary}
%\input{appendix}
\input{acknowledgements}

% bibliography
\bibliographystyle{JHEP}
\bibliography{references}

\end{document}

%% file: packages.tex
\usepackage{jinstpub} %For JINST-formatting
\usepackage{booktabs} % for much better looking tables
\usepackage{tabularx}
\usepackage{array} % for better arrays (eg matrices) in maths
\usepackage{paralist} % very flexible & customisable lists
\usepackage{verbatim} 
\usepackage{siunitx} % for easier number formatting
\usepackage{longtable} % multipage tables
\usepackage{upgreek} % needed for old arXiv hack
\usepackage{comment}
\usepackage{textcomp} % for trademark

\usepackage{url}
\makeatletter
\def\url@leostyle{%
  \@ifundefined{selectfont}{\def\UrlFont{\sf}}{\def\UrlFont{\small\ttfamily}}}
\makeatother
\usepackage{lineno} % Line numbering

%% file: header.tex
\title{Design of a mobile neutron spectrometer for the Laboratori Nazionali del Gran Sasso (LNGS)}

\input{authors}

\input{abstract}

\keywords{Detector modelling and simulations I (interaction of radiation with matter, interaction of photons with matter, interaction of hadrons with matter, etc);
Scintillators, scintillation and light emission processes (solid, gas and liquid scintillators); Particle identification methods;
Neutron detectors (cold, thermal, fast neutrons)}

\arxivnumber{2307.06660}

%% file: authors.tex
\author[a,1]{M.~Solmaz,\note{Corresponding author.}}
\author[b]{M.~Balzer,}
\author[c]{K.~Eitel,}
\author[d,e]{A.~Ferella,}
\author[f]{U.~Oberlack,}
\author[c]{U.~Pirling,}
\author[d,e]{F.~Pompa,}
\author[b]{D.~Tcherniakhovski,}
\author[c]{K.~Valerius,}
\author[b]{and S.~W\"ustling}

\affiliation[a]{Institute of Experimental Particle Physics, Karlsruhe Institute of Technology, 76021 Karlsruhe, Germany}
\affiliation[b]{Institute for Data Processing and Electronics, Karlsruhe Institute of Technology, 76344 Eggenstein-Leopoldshafen, Germany}
\affiliation[c]{Institute for Astroparticle Physics, Karlsruhe Institute of Technology,
76021 Karlsruhe, Germany}
\affiliation[d]{Department of Physics and Chemistry,
University of L'Aquila, 67100 L'Aquila, Italy}
\affiliation[e]{INFN-Laboratori
Nazionali del Gran Sasso, 67100 Assergi - L’Aquila,
Italy}
\affiliation[f]{Institute of Physics \& PRISMA$^{+}$ Excellence Cluster, Johannes Gutenberg University Mainz, 55099 Mainz, Germany}

\emailAdd{melih.solmaz@kit.edu}

%% file: abstract.tex
\abstract{
Environmental neutrons are a source of background
for rare event searches (e.g., dark matter
direct detection and neutrinoless double beta
decay experiments) taking place in
deep underground laboratories.
The overwhelming majority of these neutrons
are produced in the cavern walls by means of
intrinsic radioactivity of the rock and concrete.
Their flux and spectrum depend on time and location.
Precise knowledge of this background is
necessary to devise sufficient shielding
and veto mechanisms, improving the sensitivity of the
neutron-susceptible underground experiments.
In this report, we present the design and the
expected performance of a mobile neutron detector
for the LNGS underground laboratory. The detector is
based on capture-gated spectroscopy technique and
comprises essentially a stack of plastic scintillator
bars wrapped with gadolinium foils. The extensive simulation studies demonstrate
that the detector will be capable of
measuring ambient neutrons at low flux levels
($\sim$$10^{-6}\,\mathrm{n/cm^2/s}$)
at LNGS, where the ambient gamma flux is
by about 5 orders of magnitude larger.

}

%% file: intro.tex
\section{Introduction}
\label{sec:intro}

Underground laboratories render rare event searches possible thanks to the powerful attenuation of the
cosmic rays by rock overburden.
In this very low event rate regime,
ambient neutrons originating from the surrounding rock induce
background via capture reactions,
elastic and inelastic collisions
for
a variety of elusive searches, e.g., dark matter direct detection and
neutrinoless double beta decay experiments. It is
crucial that both flux and energy spectrum of the
environmental neutrons are measured with scrutiny
in order to evaluate their impact on the experimental
sensitivity and to help design an
adequate shield, suppressing the external neutron
background.

In deep underground laboratories,
the vast majority of the
cavern wall neutrons
are produced
in two mechanisms, the
spontaneous fission of $^{238}\mathrm{U}$
and the $(\alpha, \mathrm{n})$
reactions,
prompted by the $\alpha$-emitters
in the decay chain of
$^{238}\mathrm{U}$ and
$^{232}\mathrm{Th}$,
on the light nuclei present
in the rock, such as $\mathrm{C}$, $\mathrm{O}$,
$\mathrm{Mg}$ and
$\mathrm{Al}$~\cite{WULANDARI2004313}.
As a third mechanism, neutrons
are produced by the
muon interactions in the
rock as well.
However, the flux of muon-induced
neutrons is about two to three orders
of magnitude smaller than
that of radioactivity-induced
neutrons at large depth
underground~\cite{Mei:2005gm}.

Located 1400 m
below Gran Sasso mountains,
corresponding to
a depth of 3800
meter-water-equivalent~\cite{doi:10.1146/annurev.nucl.54.070103.181248},
Laboratori Nazionali del
Gran Sasso (LNGS) has been
hosting numerous rare event
searches. The thermal and
fast neutrons
in the laboratory
environment were measured on different
occasions
in the past~\cite{Bellotti:1985rz, Aleksan:1988qh, RINDI1988871, Belli1989, Cribier:1995zz, Arneodo:1999py, DEBICKI2009429, Haffke:2011fp, BEST20161, Boeltzig2018, DEBICKI2018133, Bruno2019}.
Based on these measurements,
projected background rates
due to ambient neutrons
with a shielding in place
were estimated
for several dark matter
and neutrinoless double beta decay
experiments at LNGS~\cite{Wulandari:2003px, Abt:2004yk, BELLINI2010169, WRIGHT201118, XENON100:2013jxx, XENON1T:2014eqx, CUORE:2017ztm}.
Ambient neutrons were likewise
surveyed in other deep underground
labs around the globe in an
attempt
to contribute to the hosted
experiments~\cite{Kim:2004wwa, Kudryavtsev:2008zzb, 2010arXiv1001.4383R, Eitel_2012, PARK2013302, Jordan:2013exa, ZENG2015108, Hu:2016vbu, 10.1093/ptep/pty133, Grieger2020, YOON2021102533}.

The results on ambient neutron fluxes
at LNGS vary considerably,
as they highly depend on the
uranium and thorium content,
and the water level of the
surrounding rock and concrete~\cite{BEST20161}.
In fact,
it was previously shown that
due to the different
compositions, the rock in
Hall C produces 10 times
more neutrons than that
in Hall A \cite{WULANDARI2004313}.
Hitherto, the neutrons were measured
at various locations
with $^3\mathrm{He}$
~\cite{Bellotti:1985rz, RINDI1988871, DEBICKI2009429, BEST20161, DEBICKI2018133} and  $\mathrm{BF}_3$~\cite{Belli1989}
counters, liquid
scintillators doped with
$^6\mathrm{Li}$~\cite{Aleksan:1988qh}
and $\mathrm{Gd}$~\cite{Bruno2019},
undoped liquid scintillator
cells interleaved by $\mathrm{Cd}$
sheets~\cite{Arneodo:1999py},
$\mathrm{NaI(Tl)}$~\cite{Haffke:2011fp}
and $\mathrm{BGO}$~\cite{Boeltzig2018}
crystal scintillators, and a neutron converter based on $\mathrm{Ca}(\mathrm{NO}_3)_2$
solution~\cite{Cribier:1995zz}.
Not only the different
measurement locations but also
the different detection
systems with
idiosyncratic systematics make
it difficult to reconcile
between the measurement
results~\cite{BEST20161}.
Detectors also covered
different neutron energy ranges.
For instance, the
best spectrum so far was
achieved with the Gd-loaded
liquid scintillator (1.2 ton)
embedded in the LVD
experiment~\cite{Aglietta:1992dy};
however,
with a high energy threshold
of \SI{5}{\MeV}~\cite{Bruno2019}.

In order to monitor the neutron
fluxes in different experimental
halls and to make the direct
comparison between these measurements
much easier, a mobile
neutron spectrometer would be
a valuable asset for the scientific
infrastructure of the LNGS. The
design of
such portable neutron counter
must meet the
following minimum requirements:
\begin{itemize}

\item Given the low neutron fluxes at $\sim$$10^{-6}\,\mathrm{n/cm^2/s}$ level, the detector
should have an effective area of 0.1-1
$\mathrm{m^2}$ to accumulate
enough statistics within a couple
of months of running, while the design should
be kept as simple as possible for transportation
and maintenance purposes.

\item Since the gamma fluxes
are about 5 orders of magnitude
bigger than the neutron fluxes~\cite{Haffke:2011fp},
an excellent gamma discrimination
is mandatory.

\item As per environmental
considerations by local authorities,
the liquid scintillators are not
currently permitted at LNGS.
\end{itemize}

Bonner Sphere Spectrometers (BSS)~\cite{BRAMBLETT1960395}
have been widely used
to measure the neutron fluxes
and energy spectra in deep underground
laboratories~\cite{Belli1989, PARK2013302, YOON2021102533, Hu:2016vbu, Jordan:2013exa, Grieger2020}. In
this method,
a set of thermal
neutron detectors (typically gaseous $^3\mathrm{He}$ or $\mathrm{BF}_3$
counters)
are placed
at the centre of moderating spheres
of different radii.
Each sphere is sensitive to
a particular neutron energy range.
The neutron energies are not
measured directly.
Rather,
the incident neutron spectrum
is inferred from the
individual count rates
in each sphere
upon an unfolding process.
Despite good gamma discrimination
and large energy range from
thermal to GeV scale,
the system provides poor energy
resolution and may be susceptible
to significant uncertainties
as a result of sophisticated
spectral unfolding~\cite{THOMAS200212}.
A detection scheme that enables
event-by-event
neutron energy reconstruction
may overcome these drawbacks,
thus being preferred over
BSS for the portable neutron
detector project.

While BSS
use passive moderators,
capture-gated neutron spectrometers~\cite{FELDMAN1991350, KAMYKOWSKI1992559, AOYAMA1993492, ABDURASHITOV2002318}
employ active moderators
in conjunction with
thermal neutron capture agents.
This technique allows
neutrons to be recorded
thanks to the coincidence
between the neutron capture
signal and the neutron-induced
proton recoil signals
in the active medium.
The sum energy of the proton recoils that
precede the capture signal,
in principle,
adds up to the
incoming neutron energy;
hence event-by-event neutron energy
reconstruction can be realized.
Since liquid scintillators are
not a viable option at LNGS,
plastic scintillators as an
active medium
in the mobile neutron counter
become an apparent choice.
Segmented capture-gated
neutron spectrometers,
combining plastic scintillators with
$^3\mathrm{He}$ proportional counters,
were already
developed to operate in underground laboratories~\cite{LANGFORD201578, Langford_2016},
representing a candidate detection
scheme for this particular
project.
However, such hybrid detector
design is not favorable
due to the detector complexity
and large size.

Table~\ref{tab:CaptureAgents}
summarizes
various properties
of common neutron capture
agents that can be utilized
in combination with plastic scintillators.
Development of lithium-loaded plastic scintillators
has recently been an active area of
research~\cite{ZAITSEVA2013747, BREUKERS201358, CHEREPY2015126, Ellis2017, Nemchenok2021}. To the best of
our knowledge, commercial
production has not been launched to date.
Plastic scintillator-based capture-gated
neutron spectrometers incorporating
$^{6}\mathrm{LiF/ZnS(Ag)}$
scintillators~\cite{Potapov2015, WILHELM2017}
and $^{6}\mathrm{Li}$-enriched
glasses~\cite{BARTCZIRR1994532, Nattress2016} have been previously
demonstrated.
Neutron capture pulses generated
in the $^{6}\mathrm{Li}$-doped medium
generally have
longer tails than the background pulses
produced via Compton scattering of
gamma rays in the plastic scintillator.
Thus,
the neutron
identification is accomplished by
an offline
pulse shape discrimination analysis.
A big drawback of this approach is that
as the
background rate induced by the ambient gamma
radiation is
many orders of magnitude larger than
the neutron-induced signal rate at LNGS,
the data acquisition system
would experience a heavy load during
months-long measurement campaigns.

\begin{table}[htbp]
\caption{Properties of candidate thermal neutron absorbing materials
for the LNGS neutron detector.
$\sigma_{\mathrm{n}}$ denotes
the thermal neutron capture cross section.
For $^{10}\mathrm{B}$ neutron captures,
only the dominant reaction branch is
shown with the corresponding branching ratio. No correlated gamma is
emitted in the other
reaction branch.}
\centering
\begin{tabular}{|c|c|c|c|}
\hline
Isotope & Natural Abundance (\%) & Reaction & $\sigma_{\mathrm{n}}$ (b) \\
\hline
$^{6}\mathrm{Li}$ & 7.5 & $^{6}\mathrm{Li}+\mathrm{n}\rightarrow$
$^3\mathrm{H}+\alpha$ & 940 \\
\hline
$^{10}\mathrm{B}$ & 19.8 & $^{10}\mathrm{B}+\mathrm{n}\rightarrow$
$^7\mathrm{Li}+\alpha+\gamma\,(\SI{0.477}{\MeV})$ (93.7\%) & 3840 \\
\hline
$^{113}\mathrm{Cd}$ & 12.2 & $^{113}\mathrm{Cd}+\mathrm{n}\rightarrow$
$^{114}\mathrm{Cd}+\text{$\gamma$-rays}\,(\SI{9.04}{\MeV})$ & 20600 \\
\hline
$^{155}\mathrm{Gd}$ & 14.8 & $^{155}\mathrm{Gd}+\mathrm{n}\rightarrow$
$^{156}\mathrm{Gd}+\text{$\gamma$-rays}\,(\SI{8.53}{\MeV})$ & 60900 \\
\hline
$^{157}\mathrm{Gd}$ & 15.6 & $^{157}\mathrm{Gd}+\mathrm{n}\rightarrow$
$^{158}\mathrm{Gd}+\text{$\gamma$-rays}\,(\SI{7.95}{\MeV})$ & 254000\\
\hline
\end{tabular}
\label{tab:CaptureAgents}
\end{table}

Capture-gated neutron spectroscopy
based upon commercially available
boron-loaded plastic scintillators
is a well-established concept~\cite{FELDMAN1991350, KAMYKOWSKI1992559, HOLM201448}.
The heavy capture products ($^{7}\mathrm{Li}$ and $\alpha$)
induce \SI{93}{\keV} electron-equivalent
($\mathrm{keV}_{\mathrm{ee}}$)
scintillation~\cite{FELDMAN1991350}, which is localized.
Additionally, \SI{477}{\keV} $\gamma$
could be fully absorbed depending
on the total scintillator size,
leading to a maximum possible capture
signal at \SI{570}{\keV}. Unfortunately,
the ambient gamma field at LNGS,
extending up to \SI{2.6}{\MeV}~\cite{Haffke:2011fp},
would outweigh the capture signals
in rate,
hampering pristine neutron identification.
Aside from that,
a boron-loaded plastic scintillator batch
of the size of our interest
is prohibitively
expensive. An alternative boron-based
design would involve
combining unloaded plastic scintillators
with gaseous boron trifluoride
($\mathrm{BF}_3$) thermal
neutron detectors. However, we already
abandoned this hybrid scheme
due to the reasons stated above.

Two isotopes, $^{155}\mathrm{Gd}$ and
$^{157}\mathrm{Gd}$, with very large neutron capture
cross sections make gadolinium highly
attractive. Despite the fact that plastic
scintillator technologies with
Gd-loading were explored by various
groups~\cite{OVECHKINA2009, Dumazert2016, Poehlmann:2018sto},
Gd-doped plastic scintillators are yet to be
commercialized. A novel type of Gd-based capture-gated
neutron detector was previously
demonstrated~\cite{PAWELCZAK2011, KURODA201241, MULMULE2018104},
in which blocks of undoped plastic scintillators
were interleaved with gadolinium sheets.
The cascade of 3-4 post-capture $\gamma$-rays
$-$whose energy sum is \SI{8.53}{\MeV} and
\SI{7.95}{\MeV} for $^{155}\mathrm{Gd}$ and
$^{157}\mathrm{Gd}$ captures, respectively$-$
leaves energy deposit
in the scintillator above the gamma background
level
with an efficiency that increases with size.
This enables neutron capture events
to be distinguished
on a quasi-background-free basis
with a sufficiently large detector at LNGS.
Furthermore, since the high light yield and
the transparency of the plastic scintillators
are not compromised by Gd-loading,
the proton recoils preceding the capture
signal can be reconstructed more accurately, hence
better neutron energy resolution. As we will show
later, this detection scheme fulfills the main
requirements of the project.
Alternatively, cadmium could be
chosen instead of gadolinium,
as presented in Ref.~\cite{Liu_2016}.
However,
the use of cadmium was avoided due to its toxicity.

%% file: design.tex
\section{Design}
\label{sec:Design}

\begin{figure}[htbp]
\centering
\includegraphics[width=.55\textwidth]{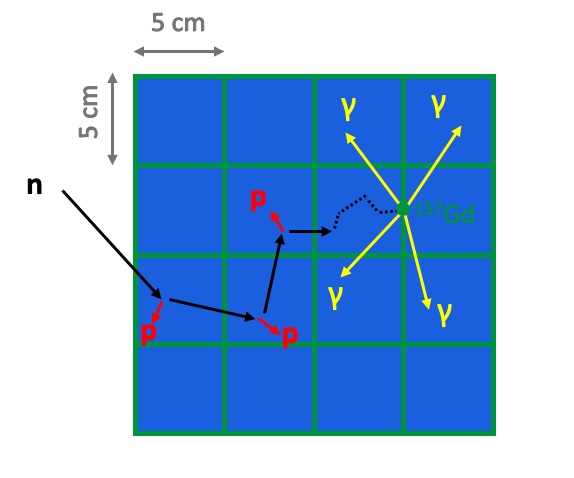}
\caption{A simplified drawing of a 
fast neutron event in the LNGS Gd-based
capture-gated neutron spectrometer.
A fast neutron (black) slows down
via proton recoils (red) in the plastic
scintillator bars (blue). The thermalized
neutron (dashed black) wanders
in the plastic and eventually gets captured
in the Gd foil (green).
Post-capture $\gamma$-rays (yellow)
serve as a neutron tag.
Preceding proton recoils permit
energy reconstruction of the
incident neutron.}
\label{fig:detPrincip}
\end{figure}

The detection principle of the LNGS mobile
neutron spectrometer is illustrated in
Figure~\ref{fig:detPrincip}. The detector essentially
consists of a stack of plastic scintillator
bars. The four sides of each bar are
covered by thin Gd foils to enhance the
neutron detection
sensitivity. A fast neutron loses
energy through proton recoils in the scintillator
bars. Following the thermalization, the neutron
is captured in the Gd foil, resulting in
emission of multiple high energy
gamma rays.
Since the most energetic gamma line
present in the natural radioactivity of
the LNGS cavern
is at \SI{2.6}{\MeV} ($^{208}\mathrm{Tl}$
decay in the $^{232}\mathrm{Th}$ series),
an energy deposit above this level would
be identified as a neutron capture signal
with a negligible background. The capture of a fast neutron
would also reveal that the neutron
lost essentially all of its
energy in the scintillator bars.
Therefore, the
neutron energy can be reconstructed based on prior
proton recoil signals. Time correlation
between  delayed capture and
preceding recoil interactions
establishes a ground to remove the
uncorrelated background pulses imitating
proton recoils. Additionally,
the capture of ambient thermal neutrons on
the support structure and gamma shield is possible and
the resulting high energy $\gamma$-rays will
be detected by the system as well. However,
thermal neutron capture signals will not be preceded
by time-correlated proton recoils.

\begin{figure}[htbp]
\centering
\includegraphics[width=.45\textwidth]
{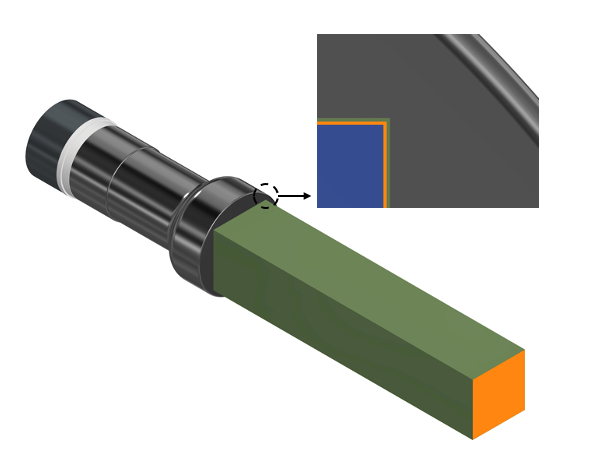}
\includegraphics[width=.5\textwidth]{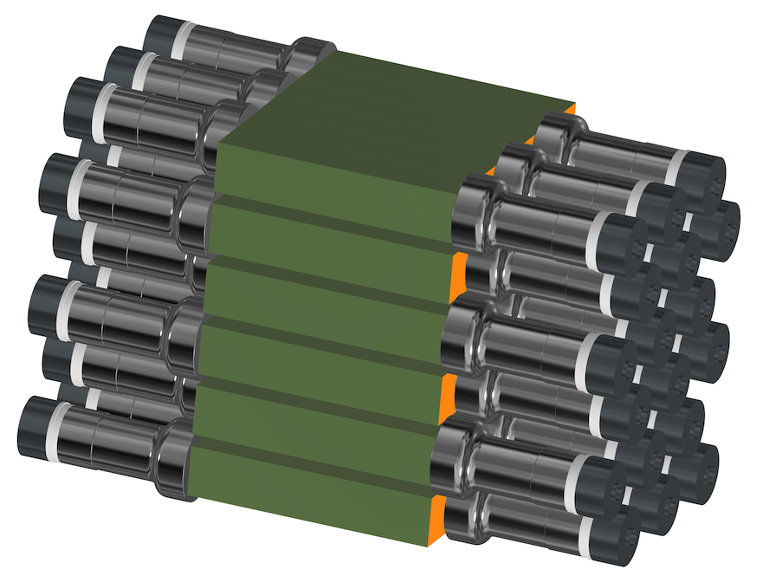}
\caption{(Left) Illustration of an individual detector
segment. A 3-inch 9302B PMT
is glued to a 5 cm x 5 cm x 25 cm EJ-200
plastic scintillator
on the head with optical cement.
The scintillator is
first wrapped by a reflector
film (orange) on remaining
five sides, and then covered
by 100 $\mu$m thick 5 cm x 25 cm
Gd foils (green) on four lateral
sides. The inset provides a
longitudinal view of the
PMT-scintillator intersection around
the corner of the scintillator (blue),
indicating that all points
on the scintillator face see the PMT.
(Right) CAD drawing of the LNGS mobile neutron spectrometer. The detector
is assembled as a 6 x 6 array of
detector segments.
A gap of 1.2 cm is present
between the scintillator rows, while
there is no horizontal gap
between the scintillator columns. To elude
the overlap between PMTs,
every two adjacent PMTs are imposed
to face opposite directions.
The support structure,
the gamma shield as well as
light-tight black covers
are not shown in the
drawings to avoid cluttering.}
\label{fig:SingleAndWholeDet}
\end{figure}

Since Gd-loading dictates a segmented
structure, the detector can be
composed of independent detector
segments. Figure~\ref{fig:SingleAndWholeDet} (left) portrays a single detector
module, which essentially consists
of a 3-inch photomultiplier
tube (PMT) and a plastic scintillator
bar of dimensions 5 cm x 5 cm x 25 cm.
The PMT is of 9302B series,
manufactured by
ET Enterprises
Limited~\cite{ET} using ultra-low
background glass.
The PMTs are equipped with mu-metal shields against
Earth's magnetic field.
The plastic scintillator is
of EJ-200 type fabricated by
Eljen Technology~\cite{Eljen}. To improve the
light collection, the scintillator
is wrapped by 3M ESR
(Enhanced Specular Reflector)~\cite{3M} film
on all sides not facing the PMT.
The scintillator is glued to
the PMT window by a thin layer of optical cement
(EJ-500 by Eljen Technology)
to accomplish optical coupling
and mechanical stability. Finally,
the reflector layer is covered by
100 $\mu$m thick Gd foils on all four lateral sides. These foils
are 5 cm by 25 cm in size and
supplied by Stanford Advanced
Materials~\cite{SAM}.
The production process of Gd foils
encompasses several stages, namely,
melting, purification, ingot casting,
extrusion, rolling, and surface treatment,
by which $\geq$99.9\% purity (in weight) is attained.
The mobile neutron spectrometer is made of 36 identical detector
modules in 6 x 6 arrangement
as shown in Figure~\ref{fig:SingleAndWholeDet} (right). Note that
all modules are individually
wrapped by thin black
foils, not displayed in the figure,
in order to ensure light-tightness.
The rows in the
scintillator array
are separated by a vertical gap of
1.2 cm, whereas there is no horizontal gap
between the scintillators within
each row.
For every pair of
neighboring detector modules,
no two PMTs are situated
on the same side of
the scintillator array.
Considering that the PMT window
(3 inch diameter) has
larger area than the
scintillator face (5 cm x 5 cm),
this detector layout achieves the best
possible compactness. In addition,
the detector architecture includes
a gamma shield, which will be discussed
in Section~\ref{subsec:Shield}.

\subsection{Determination of detector segmentation and segment size}
\label{subsec:JustDetDimensions}

There are major design constraints to impose for practical as well as physical reasons.
First of all, the number of PMT channels
should be reasonably small,
in fact not
greater than 40, which allows to use the
existing
data acquisition (DAQ) board provided
by the Institute for Data Processing and Electronics
at Karlsruhe Institute of Technology (KIT) for this
project. The details regarding the DAQ system
will be presented in Section~\ref{sec:Setup}.
Secondly, light guides are avoided
because of adding complexity and extra dead material
to the detector, and attenuating the signal size.
Thirdly, the light collection
within a detector segment should be fairly uniform,
i.e., no significant position dependence
to ensure
good energy resolution. Finally, the detector
should have an adequate size (surface area of $\sim$0.5 $\mathrm{m}^2$)
and a decent neutron efficiency ($>$3 \%) to record at least 10
neutrons (fast) per day at LNGS,
if a total neutron flux of
$10^{-6}\,\mathrm{n/cm^2/s}$ is assumed.

The uniform light collection without the aid of light guides
can be achieved using transparent but
not too long
plastic scintillator bars. If the scintillator slabs
are short enough to ensure the homogeneous
light collection, then one PMT
per plastic bar is sufficient. Given the number of channels in the DAQ board, the detector
can comprise up to 40
scintillator-PMT pairs. The scintillators
can be a cylinder or a cuboid.
Based on previous
applications and good practice,
cuboid plastic scintillators were
preferred
despite that the coupling between a PMT
and a cylindrical bar appears much
simpler\footnote{Optical
simulations show that
a cylindrical scintillator
results in larger light collection
efficiency on average yet
larger positional variations than
a cuboid scintillator of similar size.
}. Furthermore,
discarding the light guide-based
approach requires that for uniform light detection there must be no
optically-decoupled spot on the scintillator-PMT interface.
Thus, the PMT window size is a natural
guide to the transverse
dimensions of the scintillator bars.
The typical PMT choices are 2- and 3-inch diameter ones and they can
cover the entire scintillator
cross-sections of 3.5 cm x 3.5 cm and 5 cm x 5 cm, respectively.

A scintillator size of 30 cm x 30 cm x 25 cm (surface area of 0.48 $\mathrm{m}^2$)
was deemed adequately large for the
neutron flux measurements and
picked as a starting point to explore
detector segmentation.
The scintillator volume
could be cut into segments of size of 3.5 cm x 3.5 cm x 25 cm if 2-inch PMTs
were used. This would require more than 40 channels in the DAQ board,
which can not be accommodated. In the scenario of 3-inch PMTs, the same volume
could be divided into 36 segments with the dimensions of 5 cm x 5 cm x 25 cm,
satisfying the channel requirement. The
choices for the PMT window size
as well as the transverse dimensions of the scintillator segments
were then made right away.

Although the segment cross-section
was determined, the segments
could also be assembled from
smaller segments, also wrapped
by reflector and Gd foils, to
enhance the Gd-loading,
hence the neutron efficiency.
However, the more pixels
the segments have, the more
the light collection uniformity degrades
and the worse the energy resolution becomes.
In order to examine the impact of the
design parameters
on the light collection uniformity, Geant4~\cite{GEANT4:2002zbu}
simulation toolkit, version 10.06.p02, was utilized.

Two mini-segmentation cases as well as
that of no mini-segmentation were
considered to simulate
the light collection within
a detector module. Given the
benchmark segment size of
5 cm x 5 cm x 25 cm, the dimensions
of the mini-segments were
$W$ x $W$ x 25 cm, where
the width, $W$, was set to
1.25 cm, 2.5 cm and 5 cm,
corresponding to 4 by 4 and 2 by 2 mini-segmentation,
and no mini-segmentation,
respectively. Each mini-segment
was covered by a reflector
layer on all sides except the front
end. The scintillator faces
and the PMT window were
coupled by 125 $\mu$m thick
optical glue.
A tiny air gap was placed between
scintillator and reflector
to bolster total internal
reflections.
Gadolinium foils do not bear
an impact on the optical response,
so they were ignored in the
geometry for this study.
Table~\ref{tab:OptSims}
summarizes the detector
components and their optical
properties used in the light
collection simulations.
No wavelength dependence was assumed
for the optical parameters presented
therein.

\begin{table}[htbp]
\caption{Optical parameters used in the
Geant4 simulation of the light collection for
the LNGS neutron detector modules.
The scintillator surfaces were modelled
as polished. The attenuation length
of the scintillator was quoted from the
manufacturer's datasheet. The optical glue's
absorption length was inferred from
the measurements in Ref~\cite{EJ-500_abs}.
Based on the datasheet, the reflector has a nominal
reflectance greater than 98.5\%. However,
the reflectivity was set to 95\% to account
for minor imperfections at the air gap-reflector
interface. The optical photons were assumed
to undergo
specular spike reflection (i.e., perfect mirror)
on the reflector surface.}
\centering
\begin{tabular}{c|c|c|c}
\hline
Component & Refractive Index & Absorption Length (cm) & Reflectivity \\
\hline
EJ-200 Plastic Scintillator & 1.58 & 380 &  \\
PMT Borosilicate Glass & 1.49 & $\infty$ &  \\
EJ-500 Optical Cement & 1.57 & 1.65 & \\
Enhanced Specular Reflector & & & 0.95 \\ 
\hline
\end{tabular}
\label{tab:OptSims}
\end{table}

\begin{figure}[htbp]
\centering
\includegraphics[width=.45\textwidth]
{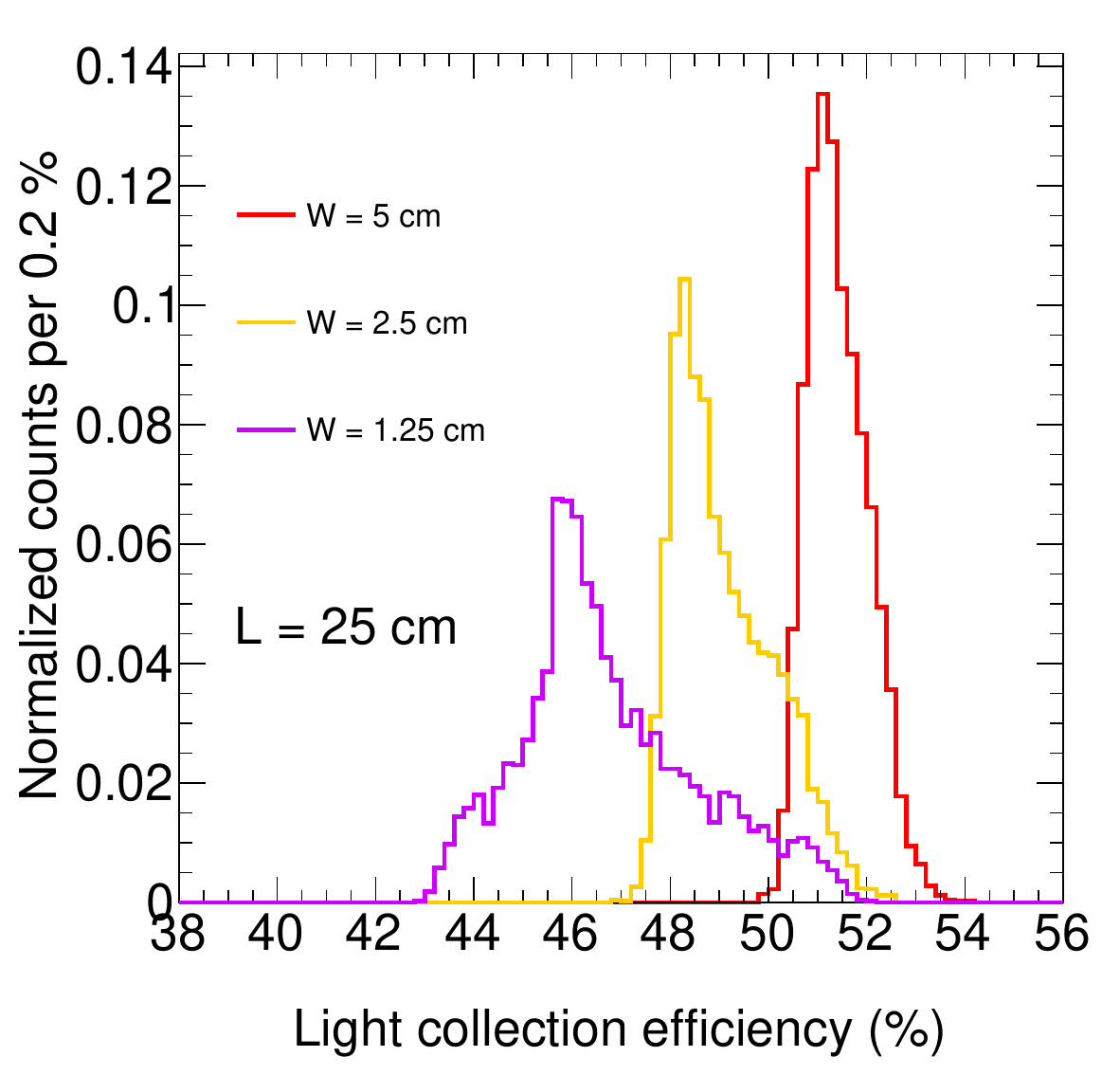}
\includegraphics[width=.45\textwidth]{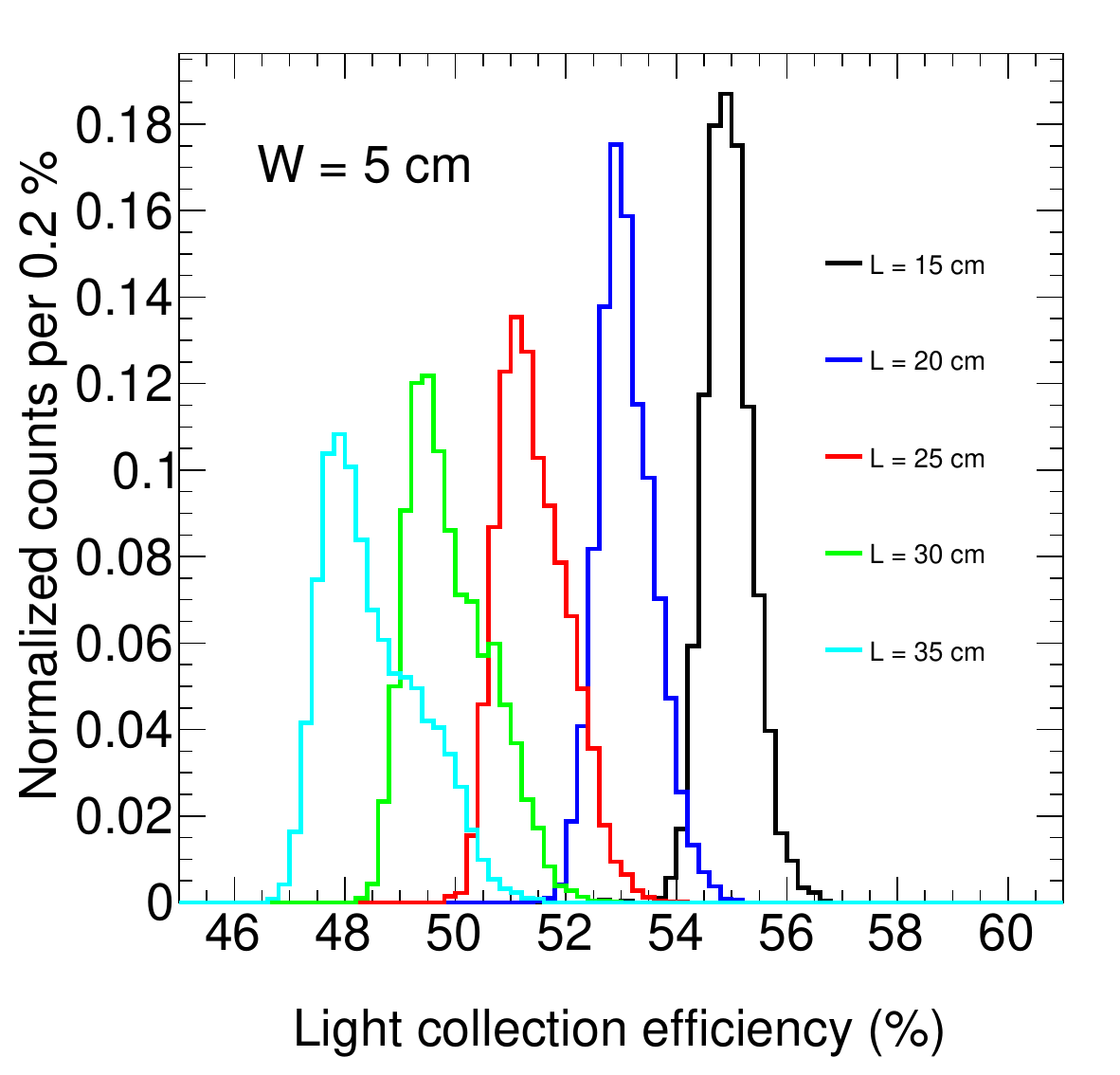}
\caption{(Left) The LCE distributions of the detector
segments, 25 cm in length,
constructed from 4x4 (violet), 2x2 (orange)
and 1x1 (red) scintillator pixels.
The histograms have mean values of 46.77\%,
49.11\% and 51.43\%, and (unbinned) standard deviations of
1.81\%, 1.01\% and 0.62\%, respectively.
(Right) The LCE distributions of the detector
segments, 5 cm in width, with different lengths.
The histograms for the lengths of 15 cm, 20 cm, 25 cm, 30 cm
and 35 cm
have mean values of 54.95\%, 53.11\%, 51.43\%,
49.91\% and 48.50\%, and (unbinned) standard deviations of
0.43\%, 0.51\%, 0.62\%, 0.74\% and 0.88\%, respectively.
All histogram contents were normalized to one.
}
\label{fig:designLCE}
\end{figure}

To compute the position-dependent light
collection efficiency (LCE),
photons were uniformly generated over the
scintillator module, which was then
divided into cells of 0.5 cm x 0.5 cm x 0.5 cm.
About 25000 photons were
originated from each cell.
The light collection
efficiency here is defined as the ratio of the
number of photons hitting the PMT photocathode
(right behind the PMT window)
to the total number photons generated
in a cell. Note that the photocathode
has an active diameter of 70 mm.
Figure~\ref{fig:designLCE} (left)
shows the light collection
efficiency distribution of the cells
for the three segmentation scenarios given
the length of the scintillator
module, $L=25$ cm. As expected, the
case without mini-segmentation
($W=5$ cm) is the most optically favorable
option. The LCE distribution for $W=5$ cm and $L=25$ cm has a mean of
51.43\% and a standard deviation of 0.62\%.
Furthermore, the impact of the scintillator length
on LCE was analyzed.
Although shorter modules
are optically more favorable,
they produce smaller neutron efficiencies,
as the efficiency is proportional to the detector size,
i.e., less likely for post-capture $\gamma$-rays
to escape without
energy deposit. By the same token, the ambient
gamma background rate also scales with the module length.
In Figure~\ref{fig:designLCE} (right), the LCE distributions
of the modules with various lengths are compared given the
width of 5 cm. The LCE distributions for $L=30$ cm and $L=35$ cm
possess mean values of 49.91\% and 48.50\% and standard
deviations of 0.74\% and 0.88\%, respectively.
Considering the trade-off between LCE, neutron efficiency
and gamma background rate, a cautious choice
of $W=5$ cm and $L=25$ cm was made for the segment size.
In Section~\ref{subsec:JustGdThickness}, it will be established
that this design provides sufficient neutron detection efficiency.

\subsection{Determination of gadolinium foil thickness}
\label{subsec:JustGdThickness}

The effect of the gadolinium foil
thickness on the neutron detection
performance was investigated within
the same Geant4 framework as in
Section~\ref{subsec:JustDetDimensions}.
The reference hadronic physics list, QGSP\_BIC\_HP,
was selected for the neutron simulations,
which includes the high-precision
transport model of neutrons with energies
below \SI{20}{MeV} based on evaluated
neutron data libraries.
At thermal energies, the neutron
wavelength becomes comparable
to the inter-atomic distances
of the hydrogenous medium,
complicating the neutron scattering
as a result
of molecular
effects~\cite{Thulliez:2021ejj}.
In order for a more accurate
modelling of thermal neutron-hydrogen
interactions (<\SI{4}{eV}) in the
plastic scintillator,
G4NeutronHPThermalScattering library
was also implemented, using the available evaluated
thermal neutron scattering data for
polyethylene as a proxy material.

Thanks to the initiative of the neutrino
physics community, it has been
widely known that
Geant4's modelling of the Gd gamma cascade
following the neutron capture has
shortcomings~\cite{GLG4Sim}.
Since many of the post-capture
$\gamma$-rays escape the detector
due to its small size,
a correct
description of the Gd deexcitation cascade in simulations
(i.e., individual gamma spectrum
and gamma multiplicity) is essential
for an accurate computation of the expected
neutron efficiency. An improved Geant4-based
Gd(n, $\gamma$) model developed by the LZ dark matter experiment was employed in neutron
simulations~\cite{LZ:2018qzl, Solmaz:2020nhm}.

In simulations, neutrons were launched uniformly
in all directions from the surface of a sphere
with a radius of 90 cm. The center of the sphere
coincides with that of the full detector assembly shown
in Figure~\ref{fig:SingleAndWholeDet} (right).
Neutrons
follow the standardized energy distribution of
a bare $^{252}\mathrm{Cf}$ source~\cite{Weinmann2017}
with a low energy cut-off at \SI{1}{MeV}.
The $^{252}\mathrm{Cf}$ spectral shape above this energy
is a good approximation for that of the LNGS ambient
neutrons calculated in Ref.~\cite{WULANDARI2004313}.
Three Gd-foil
thicknesses were investigated,
namely 10, 30 and \SI{100}{\micro\metre}.
100 million neutrons were generated for each
thickness configuration.

\begin{figure}[htbp]
\centering
\includegraphics[width=.45\textwidth]
{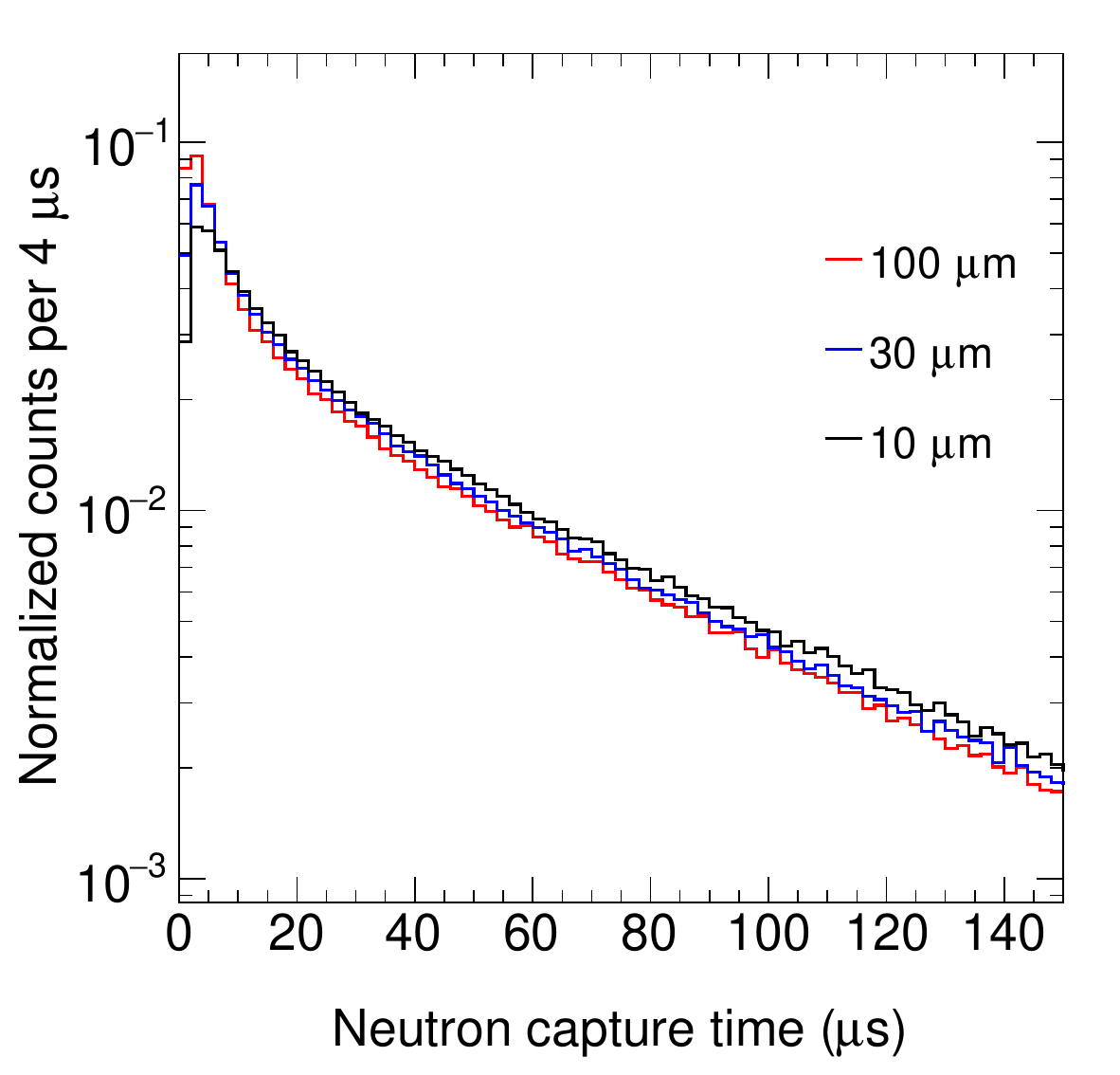}
\includegraphics[width=.45\textwidth]{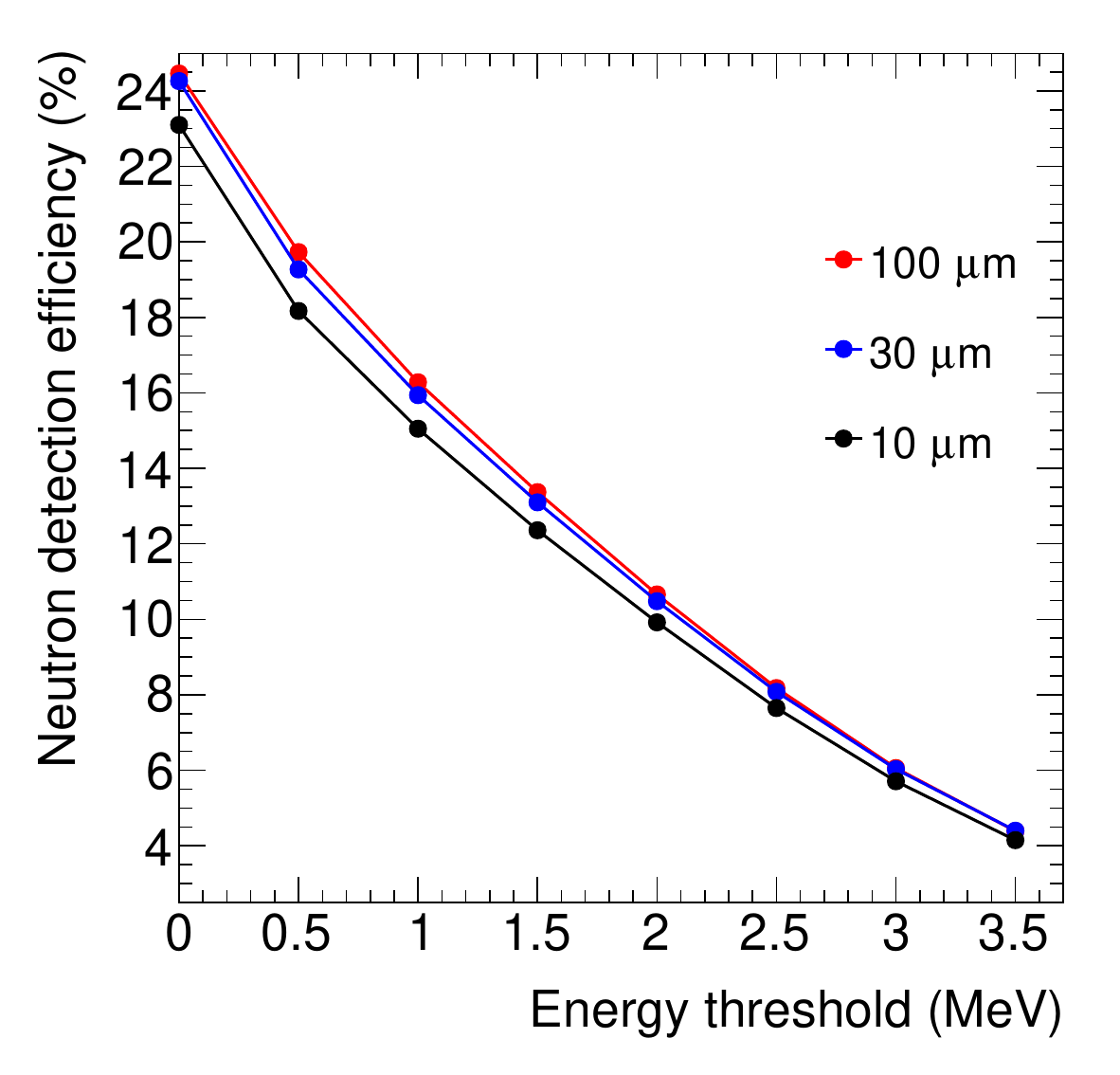}
\caption{(Left) Normalized neutron capture time
distributions for three Gd foil thicknesses.
The capture time was defined with respect
to the generation time of the neutrons.
The first proton recoil typically
takes place \SI{40}{\nano\second}
after the neutrons start to propagate.
The 95th percentile of the distributions
is at about \SI{150}{\micro\second}.
(Right) Neutron detection efficiency as a function
of the energy threshold on the post-capture gamma
deposits in the scintillator.
A capture time cut of [1-100] $\mu$s
was implemented.
The efficiencies were simulated for
the full geometry of the setup.
Only neutron captures on Gd were included
in both figures.
}
\label{fig:captureTimeEff_TrueEnergy}
\end{figure}

The neutron capture time profiles
of the detector setups with different
Gd foil thicknesses are presented in
Figure~\ref{fig:captureTimeEff_TrueEnergy} (left).
The capture time distributions
are essentially very much alike except that
more prompt captures take place
with increasing thickness. Ultimately,
\SI{30}{\micro\metre}
thickness is already sufficient to capture almost
all thermal neutrons reaching the Gd layer
given \SI{7}{\micro\metre} of mean free path
of the thermal neutrons in Gd.
However,
Gd isotopes also have appreciable capture cross sections
above thermal energies.
As the thickness increases, more neutrons
are captured rapidly
prior to the thermalization. These instant
captures may pose a challenge for the detector
electronics, since the proton recoil and neutron
capture signals are not well-separated.
Thus, we required the capture time to be greater than
\SI{1}{\micro\second} as a conservative measure
in our analysis. Furthermore,
about 88\% and 95\% of the
captures occur within \SI{100}{\micro\second}
and \SI{150}{\micro\second}, respectively.
The upper
end of the coincidence window was then set to
\SI{100}{\micro\second} in order to detect
the most of the captures, while minimizing
the gamma accidentals.

Figure~\ref{fig:captureTimeEff_TrueEnergy} (right)
shows the neutron detection efficiencies
for three detector configurations.
The efficiency here was defined
as the probability of a neutron
entering the active detector volume
from any direction
to be captured by Gd foils within
1 to \SI{100}{\micro\second}
after its generation
and then to induce gamma energy
deposits, whose sum is above the
selected threshold energy.
The efficiency curves
are almost identical at 30 and \SI{100}{\micro\meter},
whereas the efficiencies are slightly
lower when \SI{10}{\micro\meter} thick foils
are used. Since thinner Gd foils are
more complicated to fabricate, hence more
expensive, \SI{100}{\micro\meter}
thick foils were selected considering
all three thicknesses resulted in
similar performances.

The highest $\gamma$-line originating
from the natural radioactivity at the LNGS
cavern walls is at \SI{2.6}{\MeV}.
Therefore, the energy threshold to tag
the Gd capture must be above this value in order to
distinguish Gd capture $\gamma$-rays. 
The exact threshold value depends on
energy resolution, which will be
discussed in Section~\ref{sec:DetectorResponse}.
However, it can be inferred from
Figure~\ref{fig:captureTimeEff_TrueEnergy} (right)
that the neutron efficiency is expected to be
bigger than the minimum requirement of 3\%.
Note that the fractions of neutron capture events
on gadolinium, hydrogen and carbon are
82.8\%, 16.9\% and 0.3\%, respectively,
in the active medium of the detector.
The captures on hydrogen
were excluded in the analyses,
since hydrogen capture events produce single
\SI{2.2}{\MeV} $\gamma$-rays, which would
eventually be vetoed by the energy threshold cut.

\begin{figure}[htbp]
\centering
\includegraphics[width=.55\textwidth]
{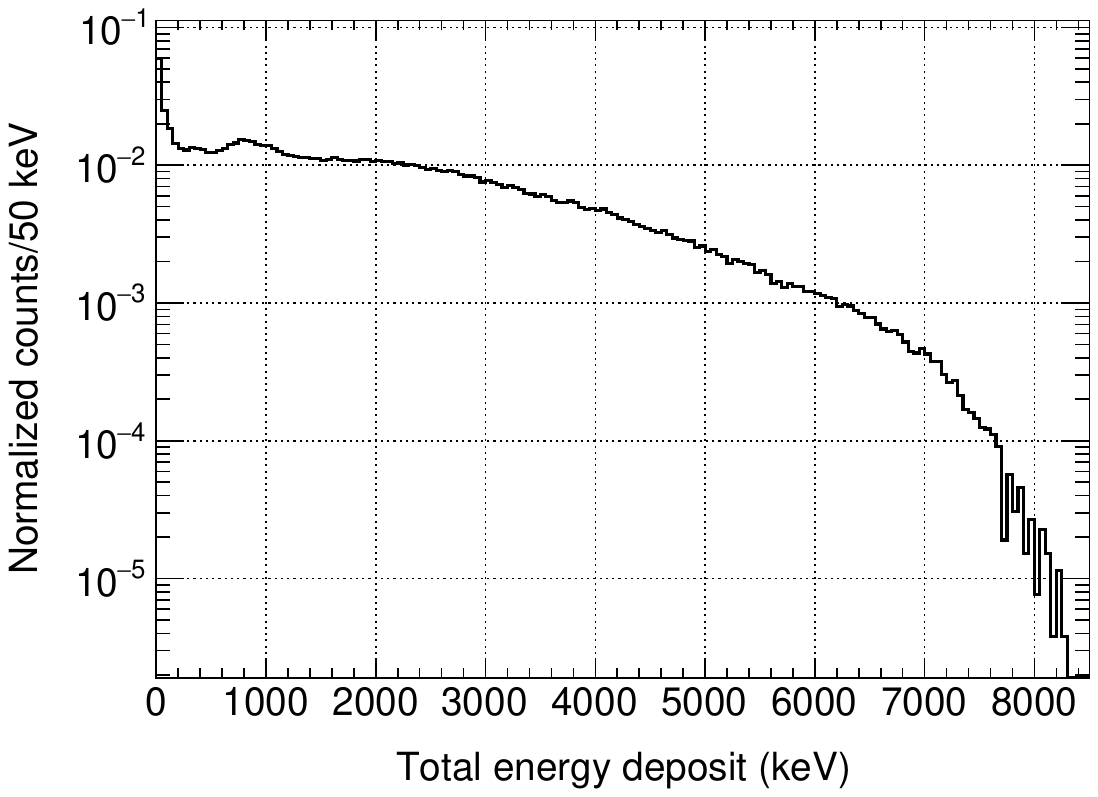}
\includegraphics[width=.4\textwidth]{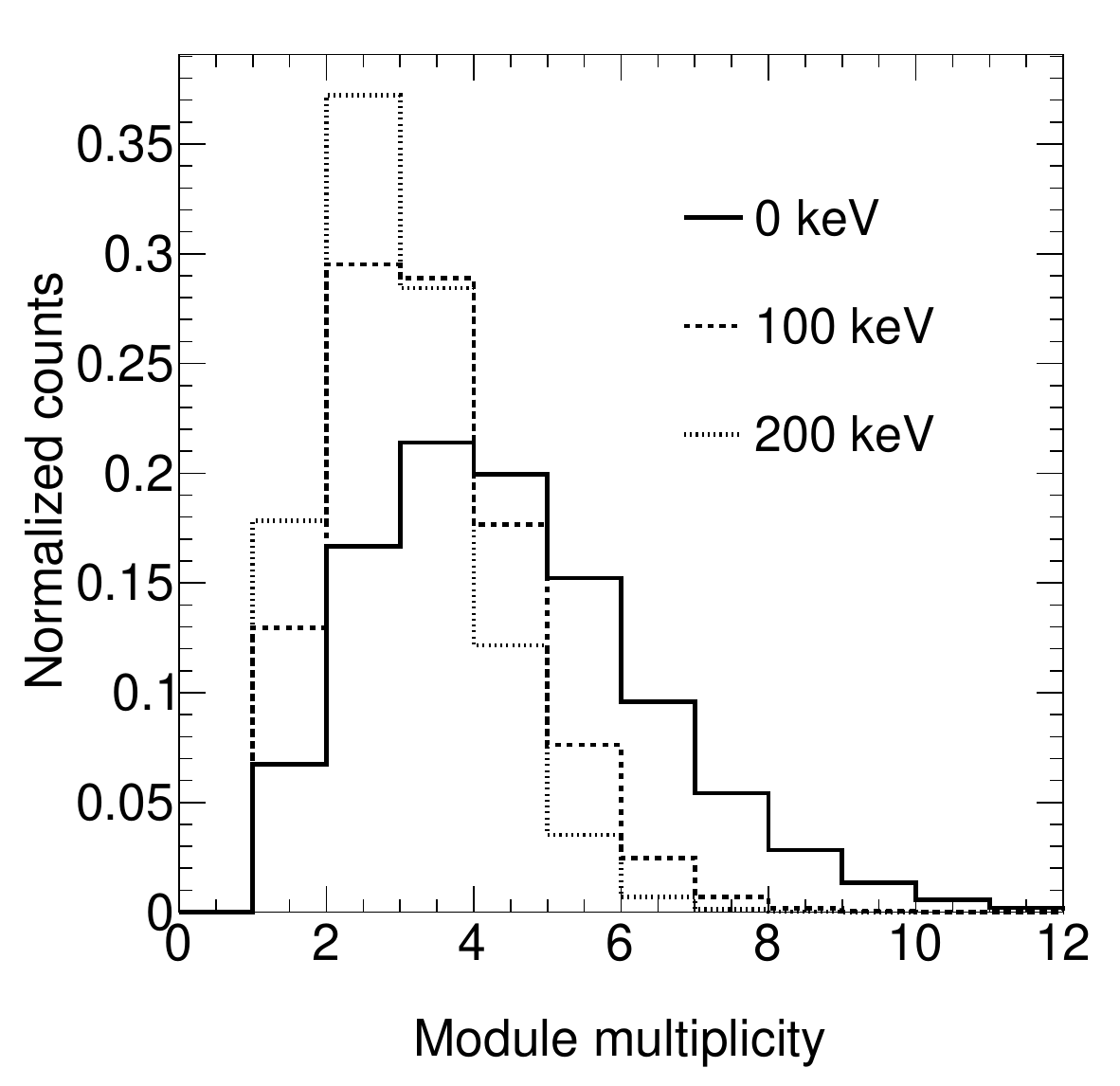}
\caption{(Left) Simulated
energy sum spectrum of the capture
events on Gd. The thickness of the
Gd foils was set to \SI{100}{\micro\meter}
in simulation.
(Right) Hit multiplicity distributions
of the Gd(n, $\gamma$) events
with the total energy deposition
greater than \SI{2.6}{\MeV}.
The multiplicities
were derived by requiring
at least \SI{0}{\keV} (solid),
\SI{100}{\keV} (dashed)
and \SI{200}{\keV} (dotted)
of energy deposition per module.
The average module multiplicities
are 3.98, 2.89 and 2.49, respectively.
All histogram contents were normalized to one.
}
\label{fig:GdEMultiplicity}
\end{figure}

Finally, we investigated the energy sum
and hit multiplicity distributions of the
Gd(n, $\gamma$) events
in order to further our understanding.
The dataset simulated with \SI{100}{\micro\meter}
thick foils was used for this purpose.
Figure~\ref{fig:GdEMultiplicity} (left)
illustrates the spectrum of the energy deposits summed
over all detector modules. Since the capture
$\gamma$-rays are not entirely contained in the detector,
the spectrum is rather featureless and ends
at about \SI{8.3}{\MeV}.
Figure~\ref{fig:GdEMultiplicity} (right)
shows the distributions of the number
of modules that are hit by the
$\gamma$-rays for the events, where
the total energy deposition is at least
\SI{2.6}{\MeV}. The distributions
were drawn for several energy thresholds
required per module. When all modules
with non-zero
energy sum are counted,
the average hit multiplicity is 3.98.

%% file: detectorresponse.tex
\section{Simulated detector response}
\label{sec:DetectorResponse}

\begin{figure}[htbp]
\centering
\includegraphics[width=.45\textwidth]
{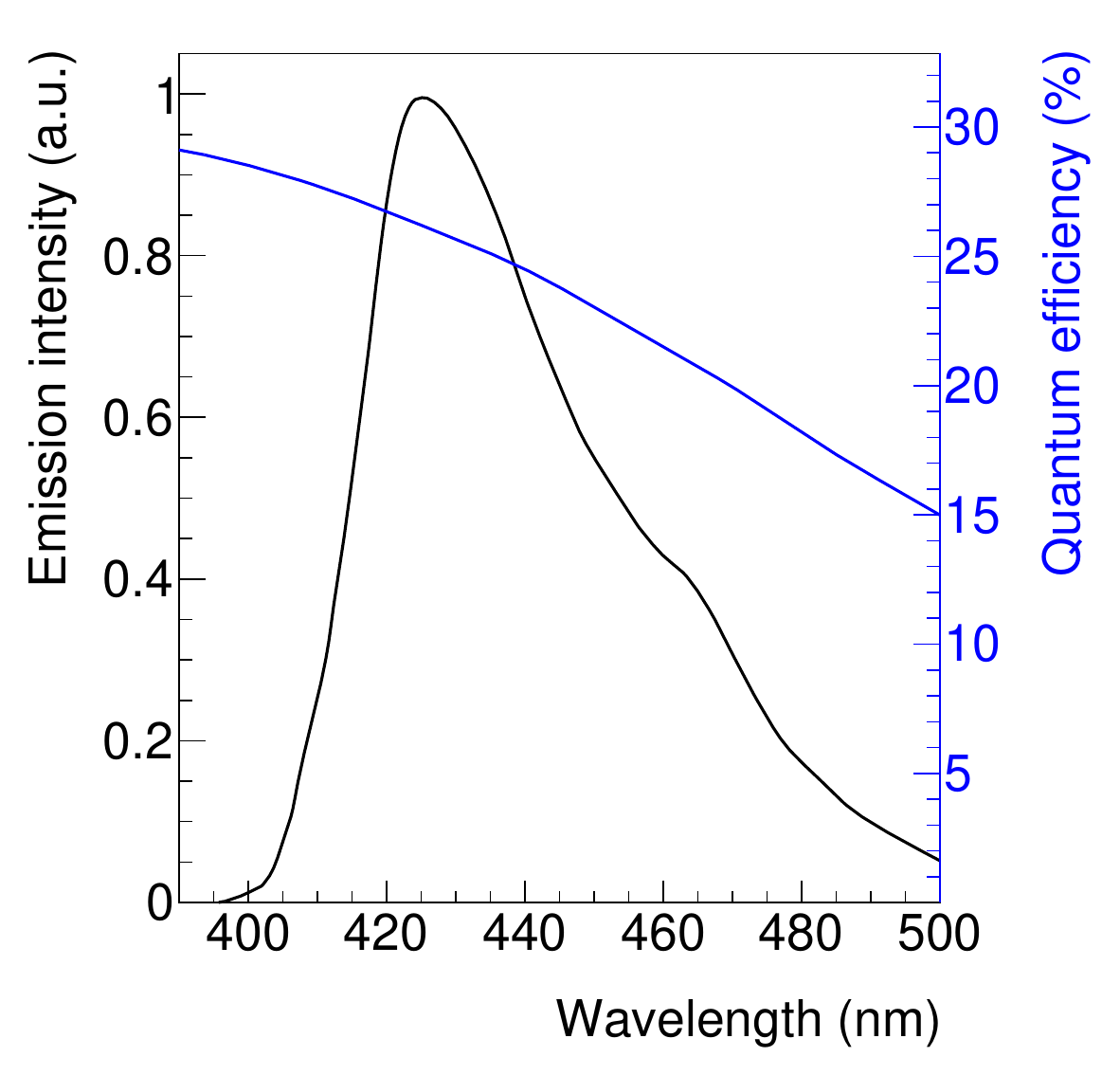}
\caption{Scintillation spectrum of the EJ-200 plastic
scintillator (black) and ET 9302B PMT
quantum efficiency vs. wavelength (blue).
Both spectra were acquired from manufacturers'
datasheets and included in the
Geant4 simulations.
}
\label{fig:scintQE}
\end{figure}

The simulation of the complete detector response to gammas
and neutrons was carried out by turning on scintillation
and quantum efficiency processes in plastic scintillators
and PMTs, respectively. Besides the optical parameters
described in Table~\ref{tab:OptSims}, the electron
scintillation yield was set to 10000 photons/MeV as per
the product specifications and the proton scintillation
yield was taken from Ref.~\cite{Laplace:2020mfy}.
Figure~\ref{fig:scintQE} illustrates the emission
spectrum of the plastic scintillator as well as
the PMT quantum efficiency (QE) as a function of the
scintillation wavelength. The effective QE averaged over
the scintillation spectrum is about 24\%. Combining it
with the mean LCE of 51.43\% and the electron scintillation
yield, an energy scale factor of 1.234 photoelectrons/keV was derived. This conversion factor was used to
obtain electron-equivalent reconstructed energy
of an event after counting the number of photoelectrons over all
detector modules.

\subsection{Suppression of
gamma-induced background}
\label{subsec:Shield}

The high ambient gamma flux at LNGS
is a potential source
of accidental coincidences
mimicking proton recoils
and neutron capture pulses in the detector.
An environmental $\gamma$-ray will be
misidentified as a real proton recoil pulse
if it randomly leaks into the coincidence
time window.
To reduce accidental $\gamma$-events
preceding a proper capture pulse,
applying a high energy threshold on the
prompt signal (proton recoil)
is not appropriate.
Plastic scintillators
yield quenched light output
for proton recoils~\cite{Birks1964}.
In EJ-200, $\sim$2.6
$\mathrm{MeV}_{\mathrm{ee}}$ signal
is produced if a \SI{6}{\MeV} neutron
deposits all its energy in a
single scatter~\cite{Langford_2016}.
Implementing a high energy cut on the
proton recoil pulse
similar to that on the
capture pulse would render
the detector
completely insensitive
to <\SI{6}{\MeV} neutrons.
Therefore, a gamma shield
should be put in place in order to
reduce
the probability of accidental
gammas mimicking a proton recoil pulse.

In order to assess the impact of the shield,
a series of gamma background simulations
was performed. The shield,
fully encapsulating the detector,
was modelled
as a hollow box with varying thicknesses.
The inner dimensions of the shield are
36.2 cm x 41.2 cm x 59 cm. Lead was chosen
as a shielding material.
The environmental gamma flux at LNGS was
measured by various groups at different
lab locations~\cite{Haffke:2011fp, Bellini:2009zw,Bucci2008,ARPESELLA1992420,Malczewski2013}. The flux numbers
range from 0.15 to 1
$\gamma$$\mathrm{/cm^2/s}$. Conservatively,
1 $\gamma$$\mathrm{/cm^2/s}$
was assumed for the gamma rate
estimation. The measured gamma spectrum
in Ref.~\cite{Bellini:2009zw,Bucci2008},
which extends up to \SI{3}{\MeV},
was fed into the Geant4 particle
generator. Taking a similar
approach as in Section~\ref{subsec:JustGdThickness},
the gamma particles were generated
isotropically from the surface of
a sphere with a radius of 90 cm.
The energy of an event was reconstructed
from the total number
of photoelectrons as aforementioned.
The gamma rates as a function
of the shield thickness were then derived.

\begin{figure}[htbp]
\centering
\includegraphics[width=.55\textwidth]
{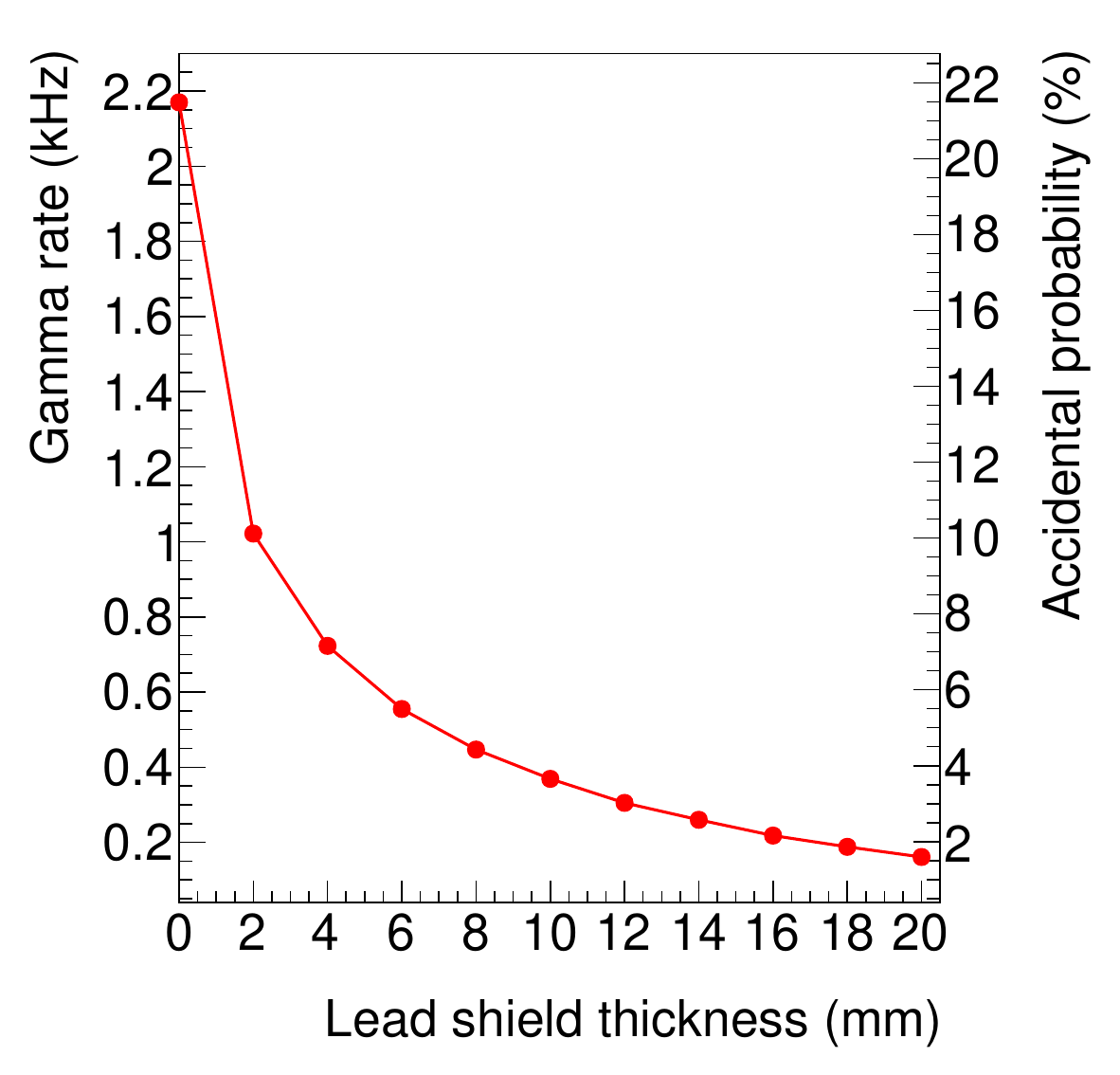}
\caption{Expected background gamma rate as
a function of lead shield thickness.
Only reconstructed energy deposits greater than
\SI{50}{\keV} are considered
in the rate calculation.
}
\label{fig:shieldImpact}
\end{figure}

Figure~\ref{fig:shieldImpact} shows the
rate of ambient gamma events leaving more than a total of \SI{50}{\keV} energy
deposit in the whole detector
as a function of the lead shield thickness.
The chosen value of \SI{50}{\keV}
for the threshold was driven by
the response of the EJ-200 scintillator to
\SI{1}{\MeV} neutrons,
corresponding to the start point energy
of the simulated neutron spectrum
in Section~\ref{subsec:JustGdThickness}.
A \SI{1}{\MeV} neutron induces on average
a proton recoil of \SI{0.5}{\MeV} energy
during a single scatter.
A proton recoil of that energy produces
slightly above 50
$\mathrm{keV}_{\mathrm{ee}}$ scintillation
in the plastic~\cite{Laplace:2020mfy}.
Since neutrons
undergo multiple
elastic collisions before the
capture,
it can be anticipated
that \SI{1}{\MeV} neutrons will induce a total
of at least
50 $\mathrm{keV}_{\mathrm{ee}}$ energy in
proton recoils most of the time.
Therefore,
only the ambient gamma events with more than
\SI{50}{\keV} energy deposit,
now a well-motivated design threshold,
could then fake the proton recoils
in our setup.
Note that this threshold energy
is equivalent to
$\sim$60 detected photoelectrons,
clearly not a weak signal.
In the real application, however, this
threshold may be revised and slightly
increased depending on noise terms originating
from the data acquisition system.
Figure~\ref{fig:shieldImpact}
indicates that the accidental probability
within the
coincidence capture time window
of [1-100] $\mu$s
can be kept
below 5\% with a 6 mm thick lead shield. Note that the
thicknesses above \SI{20}{\mm} were not
considered due to the heavy weight
compromising the mobility of the system.

External $\gamma$-rays can also induce a neutron capture
signal. In principle, these fake capture pulses
can be very efficiently
eliminated with an appropriate energy cut
given that the flux of the environmental
$\gamma$-rays with energies
above \SI{3}{\MeV}
is suppressed by five orders of magnitude
with respect to that of 
\SI{2.6}{\MeV} $\gamma$-rays
from $^{208}\mathrm{Tl}$ decays
in the cavern walls~\cite{Bellini:2009zw}.
However, two gamma rays
that randomly enter the detector
at about the same time
may produce coincidentally
summed events~\cite{Bemmerer:2005wu}.
These
2-fold pile-up events result in pulses
with energies extending up to \SI{5.2}{\MeV},
which can not be removed trivially.
This pile-up effect was exploited to estimate
the high energy background for the
capture pulses and to devise shielding
against the spurious capture events of
pile-up origin.

To incorporate the effect of the pile-up
in the background gamma spectrum, a simple model
presented in Ref.~\cite{Ahsan2021} was followed.
Suppose $f_S(E)$ represents the probability
density function (PDF) of the no-pile-up (single)
spectrum. Then, the PDF of the pile-up (double) spectrum, $f_D(E)$, 
will just be the convolution of the two single
PDFs:

\begin{equation}
f_D(E) = \int_{-\infty}^{+\infty} f_S(E^\prime)f_S(E-E^\prime)dE^\prime
\label{eqn:convol}
\end{equation}

\begin{figure}[htbp]
\centering
\includegraphics[width=.55\textwidth]
{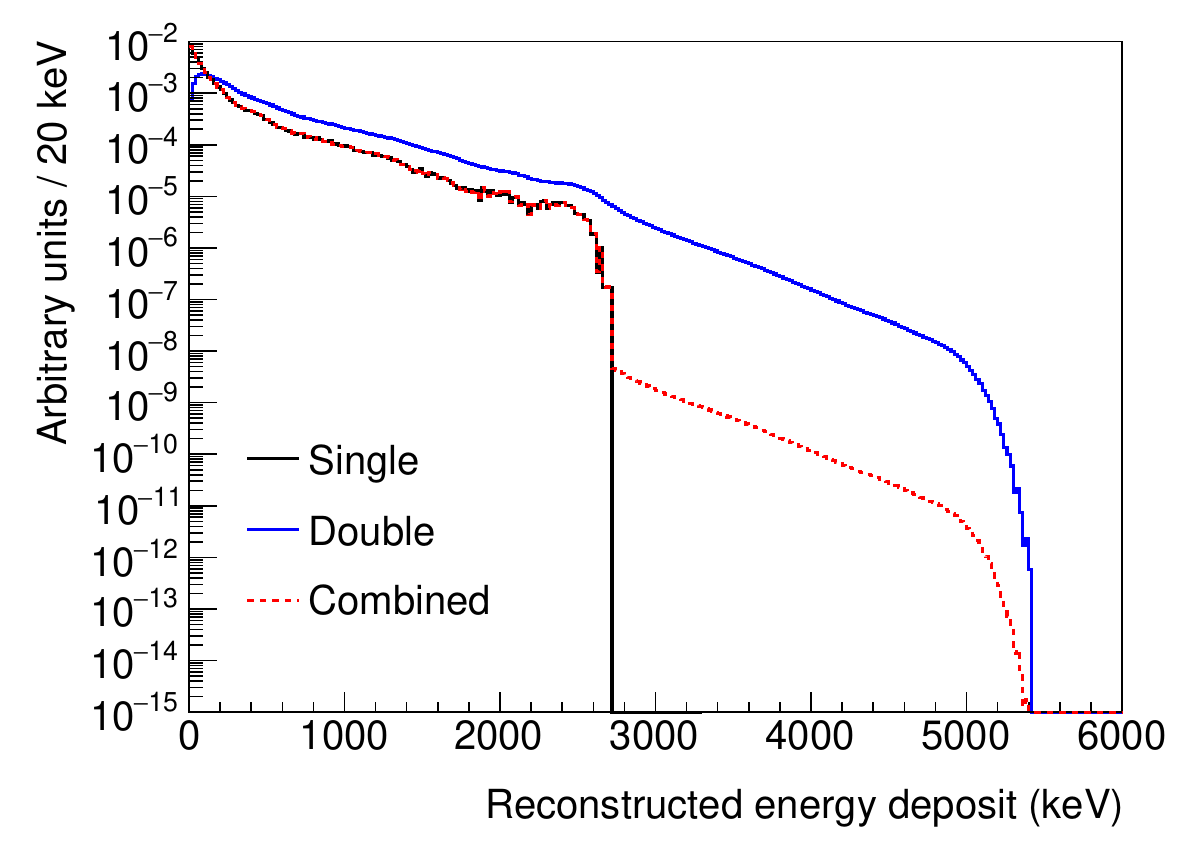}
\caption{Spectral shapes of the no-pile-up (black),
pile-up (blue) and combined (dashed red) cases
computed under the no-shield assumption.
The pile-up phenomenon is responsible for
the background above \SI{3}{\MeV}.
The high energy tail is
suppressed by the pile-up probability.}
\label{fig:pileUpPDF}
\end{figure}

Then, the combined PDF will read as:

\begin{equation}
f_C(E) = (1-p)f_S(E) + pf_D(E)
\label{eqn:combined}
\end{equation}

where $p$ is the pile-up probability. It is equal
to the single rate ($\mu$) times the pile-up time
window ($\epsilon$). A typical PMT pulse is a few tens of
nanoseconds wide. However,
the raw PMT pulses need to be stretched
due to the characteristics of
the data acquisition system (see Section~\ref{sec:Setup}).
The stretched pulses will not be wider than
\SI{250}{\ns} as a design criterion.
Thus, $\epsilon$ was set to \SI{250}{\ns}
as the pile-up time window
for the
high energy
background estimation.
Since
$p \ll 0.1$, pile-up at higher orders is negligible.
Figure~\ref{fig:pileUpPDF} provides an illustrative
example on how pile-up induces background for our
capture signals.
It should be noted that
thanks to the segmented detector
layout, two random gammas hitting different
detector segments are expected to be
resolved within a much shorter window in reality
than the assumed $\epsilon$.

The impact of the lead shield on the
pile-up-induced background
was investigated as in what follows:
First, the combined PDFs were
derived for various lead shield thicknesses.
Second, the PDFs were
normalized to the total single rate
in order to estimate high energy background
at different thresholds.
The high energy pulses alone do
not mimic a fast neutron capture event,
they have to be accompanied by
a preceding accidental $\gamma$-ray.
Thus, the rate of fake neutron captures
owing to the pile-up was obtained
after multiplying the high energy gamma
rate by the accidental probability
at a given lead thickness.

\begin{figure}[htbp]
\centering
\includegraphics[width=.49\textwidth]
{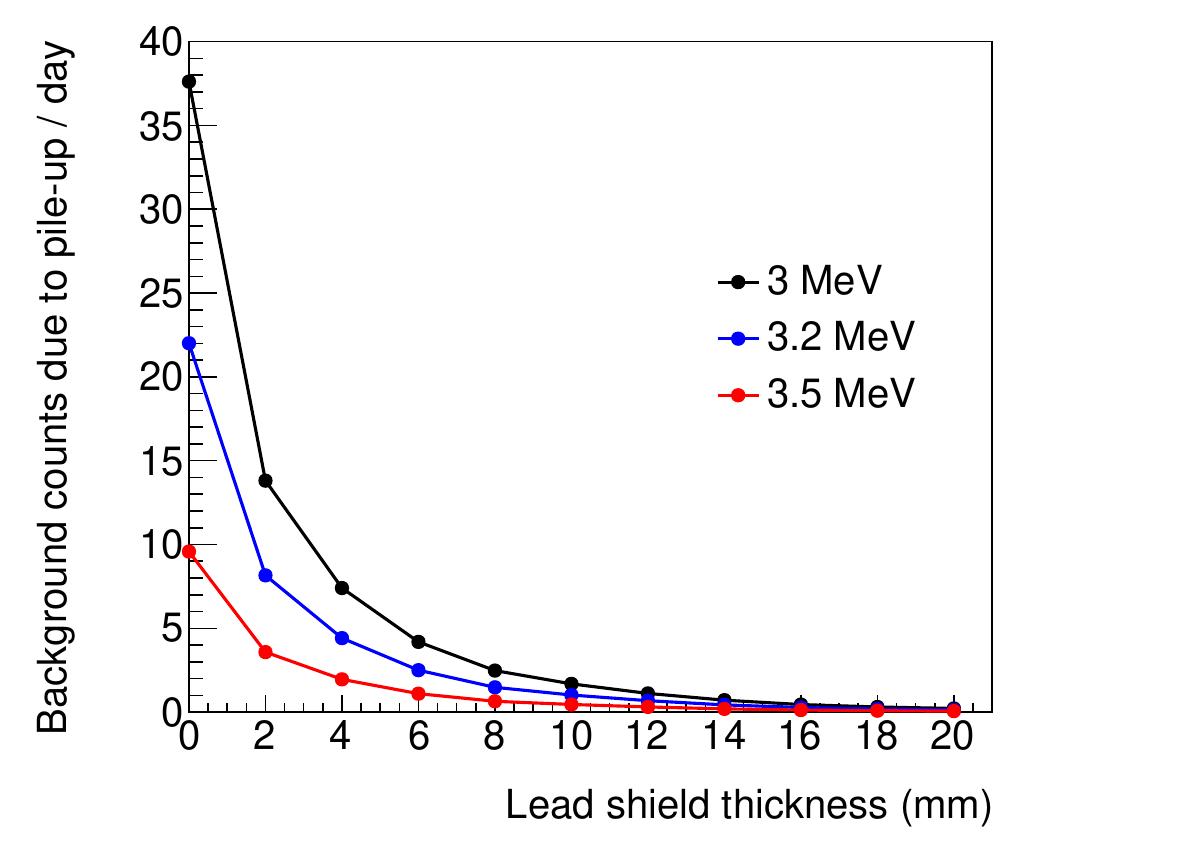}
\includegraphics[width=.49\textwidth]{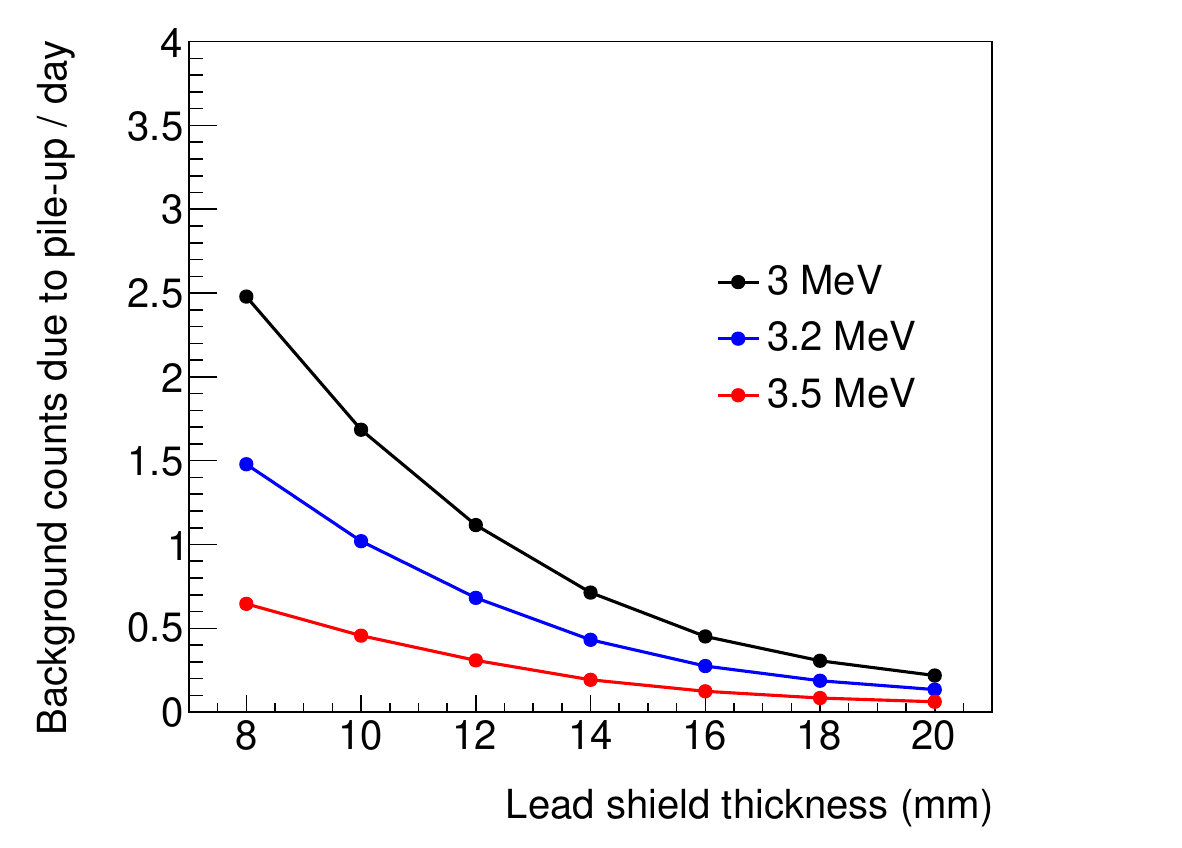}
\caption{The counts of the pile-up-induced
background per day as a function of
the lead shield thickness for three alternative
capture energy thresholds.
The right-hand panel shows the background
rates with the lead shield thicknesses
from 8 to 20 mm for clarity.
}
\label{fig:leadShieldPileUp}
\end{figure}

\begin{figure}[htbp]
\centering
\includegraphics[width=.65\textwidth]
{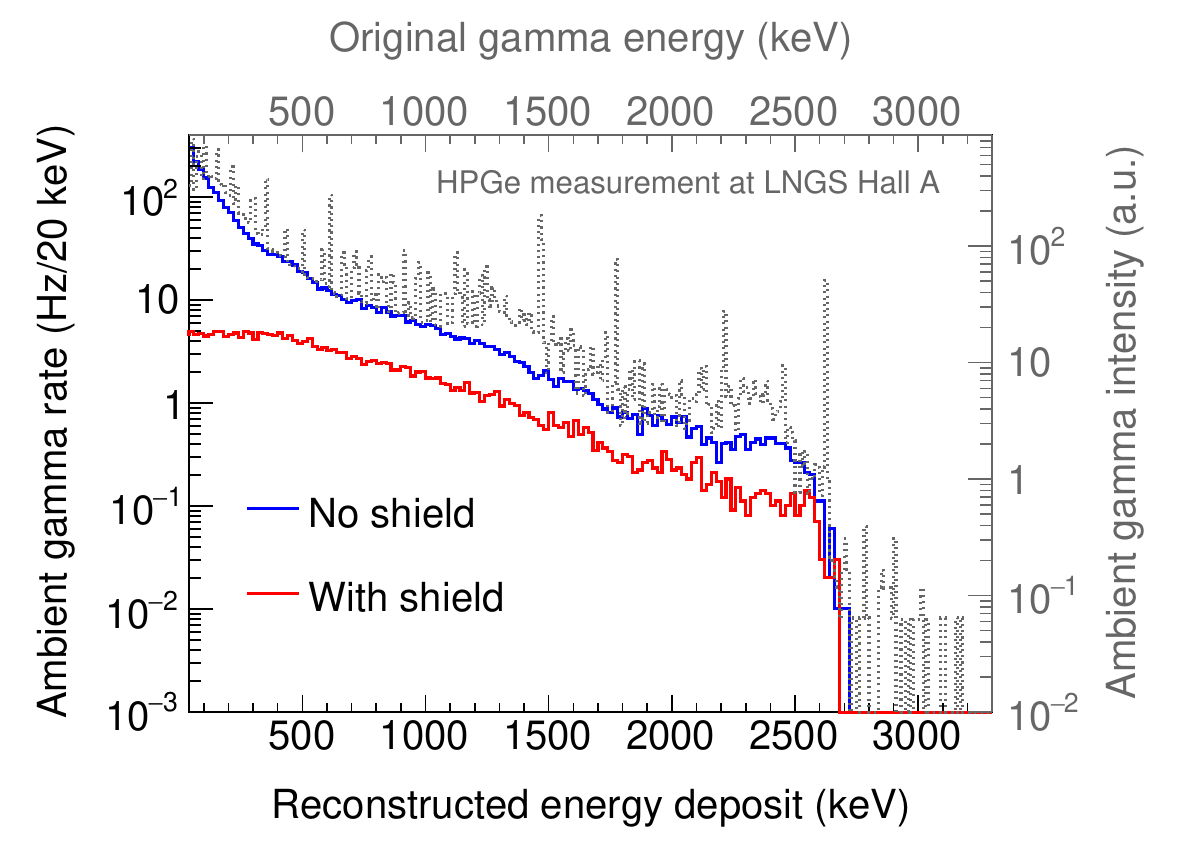}
\caption{Comparison of the
expected background gamma spectra with and
without the lead shield. The gray histogram
represents the gamma spectrum
measured at LNGS Hall A with a high-purity
germanium detector
in Ref.~\cite{Bellini:2009zw,Bucci2008},
specified as the input spectrum for the
background gamma simulations.
The integrated rates for the unshielded (blue
histogram)
and shielded (red histogram)
setups are 3 kHz and 0.227 kHz,
respectively.
A gamma flux of
1 $\gamma$$\mathrm{/cm^2/s}$
was assumed.
}
\label{fig:leadShieldBckgSpec}
\end{figure}

Figure~\ref{fig:leadShieldPileUp}
shows the rate of this background
as a function of the lead shield thickness
at three different high energy thresholds.
The background rates become almost
flat starting at 16 mm. At this thickness,
0.45 background events per day
is expected at the threshold
of \SI{3}{\MeV}, whereas
the real fast neutron expectation is
typically
10 events per day, as aforementioned.
It can be argued that
a thinner lead shield thickness may
indeed be
sufficient. However, it should be noted that
our neutron detection efficiency
is yet to be experimentally verified
and that the neutron flux at LNGS is likely
to vary significantly. The expected signal
rates may turn out to be lower or higher.
For this reason,
the lead thickness was chosen to be 16 mm,
which is rather conservative.
Note that
the accidental probability at this thickness
is 2.16\%.
Figure~\ref{fig:leadShieldBckgSpec}
presents the expected background
gamma spectra with and without the lead shield.
The background gamma rate
is suppressed by almost a factor of 15
with the lead shield.
 
\subsection{Complete detector response
to neutrons}
\label{subsec:NeutronResponse}

\begin{figure}[htbp]
\centering
\includegraphics[width=.65\textwidth]
{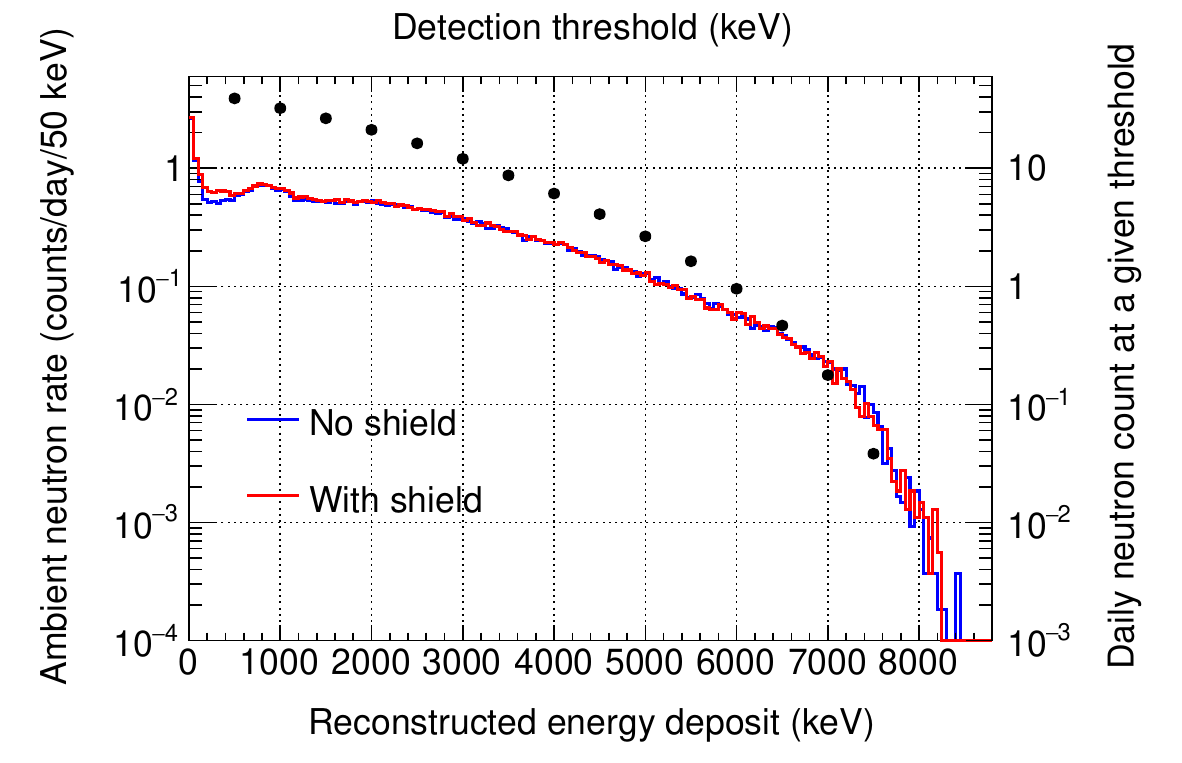}
\caption{Simulated
$\mathrm{Gd}(\mathrm{n}, \gamma)$
spectra
of the ambient neutrons (left axis)
with (red) and without (blue) a lead shield of
\SI{16}{\mm} thickness.
The black dots represent the expected
daily neutron capture counts (right axis)
at various thresholds on the
energy deposit due to capture $\gamma$-rays.
The shielded spectrum was integrated
starting at respective threshold values
in order to obtain the daily counts.
A neutron flux of
$0.42 \times 10^{-6}\,\mathrm{n/cm^2/s}$
was assumed.
}
\label{fig:captureSpectrum}
\end{figure}

The full detector response to the fast neutrons
was implemented by repeating the
simulation
steps described in Section~\ref{subsec:JustGdThickness}
with the addition of the particle type-dependent
(proton and electron) scintillation
emission. The neutron flux for the full detector
simulation was assumed to be
$0.42 \times 10^{-6}\,\mathrm{n/cm^2/s}$.
The number was based on a
previous on-site measurement
in Ref.~\cite{Arneodo:1999py} and corresponds
to the flux of neutrons with energies
between 1 and 10 MeV. In the neutron simulations,
the neutron energies are sampled from the
$^{252}\mathrm{Cf}$ spectrum,
which also has an artificial low-energy
cut-off at 1 MeV (reason explained in Section~\ref{subsec:JustGdThickness})
and an end point at 10 MeV.
The simulated gamma spectra,
following the neutron capture on Gd,
are shown in
Figure~\ref{fig:captureSpectrum}
for the unshielded and shielded
detector configurations.
The integrated neutron capture
rate is about 12 neutrons per day
at the capture threshold of \SI{3}{\MeV}.
The shield
does not affect the neutron detection
efficiency. The rate of the neutron captures
on lead leaving at least \SI{3}{\MeV}
of energy deposit in the scintillator
modules is 0.006 counts per day.

\begin{figure}[htbp]
\centering
\includegraphics[width=.65\textwidth]
{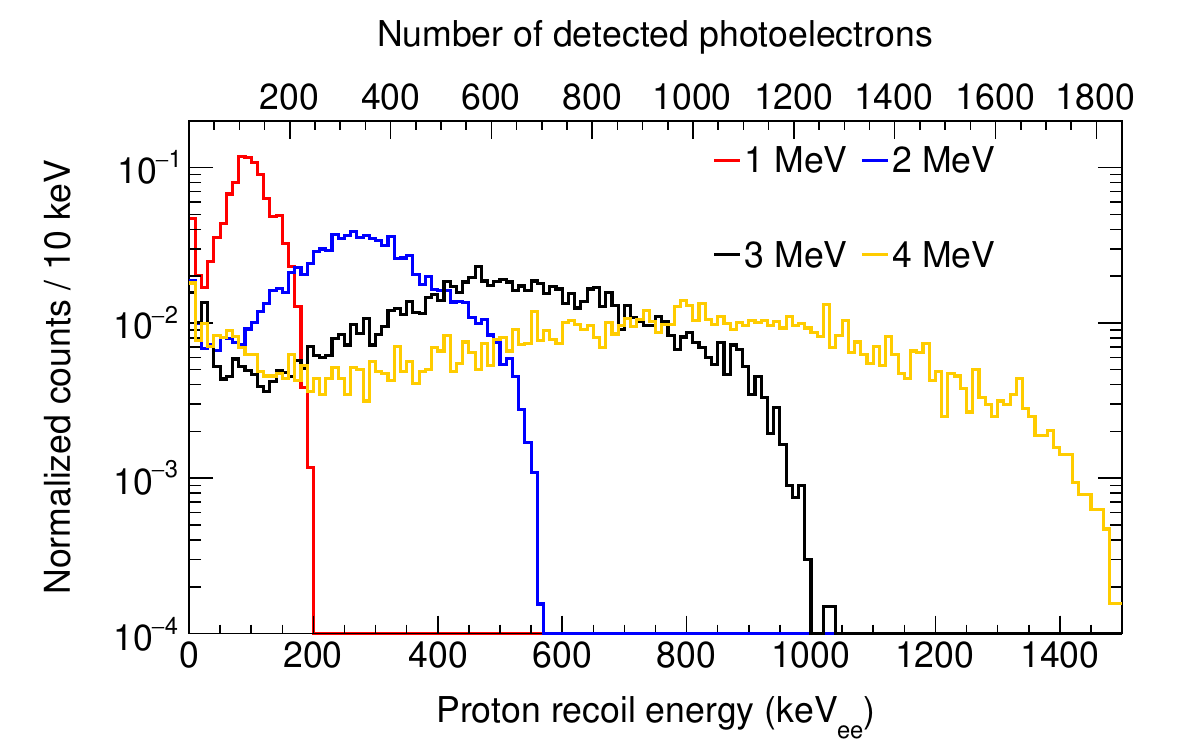}
\caption{Proton recoil spectra of various
neutron groups. Only the events
that passed the capture
time (1-\SI{100}{\micro\second}) and capture
energy (>\SI{3}{\MeV}) cuts were plotted.
The neutrons were generated
isotropically from a sphere of
a radius of 90 cm similar to the
other neutron simulations presented.
The proton scintillation was simulated
based on the measurements in
Ref.~\cite{Laplace:2020mfy}.
The detected photoelectrons
induced by proton scintillation were
counted over all channels and then
converted into an electron-equivalent
energy.
All histogram contents were normalized to one.
}
\label{fig:protonRecoil}
\end{figure}

Capture signals of the fast
neutrons will always be preceded
by proton recoils. In order to
examine our experimental sensitivity
to the proton recoils, captures
of 1, 2, 3 and \SI{4}{\MeV} neutrons
were analyzed. Figure~\ref{fig:protonRecoil}
shows the electron-equivalent energy
distributions of
the proton recoils of these neutrons.
With a probability of 85\%, \SI{1}{\MeV}
neutrons produce recoil signals,
whose sum energy is greater than the
threshold of 50
$\mathrm{keV}_{\mathrm{ee}}$.
The same probability is about 95\%
for the other three neutron groups. Thus,
operating the detector at the neutron energy
threshold of \SI{1}{\MeV} seems quite feasible.
As stated in Section~\ref{sec:intro},
Ref.~\cite{Bruno2019} reported
ambient neutron measurement at LNGS
with the best resolution up to date,
albeit at a high neutron threshold
of \SI{5}{\MeV}. By design, this neutron spectrometer
has a promising potential to reduce the
neutron threshold down to \SI{1}{\MeV} or even lower.
Note that the unfolding of the proton recoil
spectrum
and the impact of the
detector segmentation on the
neutron energy reconstruction were not
scrutinized here, as they are not within
the scope of this report.

Lastly, thermal neutrons were studied as
a background source. Thermal neutrons
can also be captured in the detector. If the
capture pulse is preceded by an accidental
gamma within the capture time window,
it would pose as a signal event.
The thermal neutron flux numbers
vary between
0.32~\cite{BEST20161}
and
5.3~\cite{Bellotti:1985rz}
$\times 10^{-6}\,\mathrm{n/cm^2/s}$
across the measurements.
Assuming the largest flux value,
the rate of thermal neutron captures on both
gadolinium and lead, leaving at least
\SI{3}{\MeV} of energy in the scintillator
volume and being coincident with an accidental
$\gamma$-ray,
is 2.26 counts per day.
In the real application,
the sum rates of this background
and pile-up-induced background will be determined
from the sideband measurements (i.e.,
studying the events where the preceding
pulse lies outside the chosen
coincidence time window),
as they are tied to random coincidences.

Muon-induced neutron activity is another
critical background source for
rare event searches at LNGS. In principle,
the designed detector
will be sensitive to neutrons of this origin
as well.
However, the flux of the muon-induced neutrons, heavily suppressed at these depths, is at least two orders of
magnitude lower than that of
radioactivity-induced neutrons, as
stated previously in Section~\ref{sec:intro}.
The event rate will therefore be statistically insignificant.
Moreover, the muon-induced neutron spectrum extends
to several GeV~\cite{Mei:2005gm}.
The design was optimized to detect
$<$\SI{10}{\MeV} neutrons, hence only muon-induced neutrons within this energy region
will be detected efficiently.
Therefore, although we may occasionally observe a muon-induced neutron event, the proposed
system is focused on the detection of
radioactivity-induced neutron activity.

%% file: setup.tex
\section{Data acquisition and mechanical setup}
\label{sec:Setup}

The data acquisition (DAQ) will be handled by the board that
was previously built at KIT for the TRISTAN
experiment~\cite{Tristan}.
The main board allows a
digital readout of up to 40 channels thanks to the
five 8-channel analog-to-digital converters (ADC)
located on individual mezzanine boards. Each ADC
board comes with an 8-channel programmable-gain amplifier
and anti-alias filters. The PMT signals are digitized
at a sampling frequency of 62.5 MHz (i.e., 16 ns of pulse
sampling width). Since the typical PMT pulses are much faster,
each raw pulse will be stretched via a low-pass circuit
individually in order for a commensurate
waveform sampling by the DAQ board. This front-end
circuitry is currently being designed and the stretched
pulses are expected be shorter than
\SI{250}{\ns} (See Section~\ref{subsec:Shield}).

The trigger decision will be made based on the sum of the
pulses from multiple detector modules that are hit
simultaneously by the capture gamma rays. This master
trigger will have a high threshold
(3 $\mathrm{MeV}_{\mathrm{ee}}$) to select
neutron capture candidate events.
Multiple modules,
where the total energy deposition is
greater than 3 $\mathrm{MeV}_{\mathrm{ee}}$,
as well as a single module
with $>$3 $\mathrm{MeV}_{\mathrm{ee}}$
energy deposit can activate the master trigger.
An event will
be centered around the master trigger time and
extend \SI{100}{\micro\second} backward and
forward.
The Field Programmable Gate Array (FPGA)
on the main board will be set up to look at all channels and
find pulses above a certain threshold
within the event window. This secondary threshold
ensures that only
waveform snippets that are significantly above the baseline
are recorded.
The coincidence time between the capture
pulse and proton recoils is not a part
of hardware trigger, but will be imposed
during offline analysis.
Note that the acquisition scheme will
grant dead time-free readout.
Additionally, the system will
be capable of adopting an external trigger as master trigger.

Ideally, a fast neutron capture
event will be triggered by the
Gd(n, $\gamma$) pulse passing the
assumed threshold of 3
$\mathrm{MeV}_{\mathrm{ee}}$, and then
preceding proton recoils with an energy sum of
at least 50 $\mathrm{keV}_{\mathrm{ee}}$
will be identified in the pre-trigger window
(\SI{100}{\micro\second} wide). However, that may not
always be the case. For instance,
a \SI{9}{\MeV} neutron,
captured in the detector, may produce
$>$3 $\mathrm{MeV}_{\mathrm{ee}}$ proton recoils,
which would trigger an event.
In this case, the actual capture pulse would follow
the trigger pulse and could be found in
the post-trigger window
(also \SI{100}{\micro\second} wide).
Thus, the event would be constructed
in the correct order
during offline analysis and the proton recoils would be selected (with no high-energy cut-off)
if and only if the capture pulse passes
the relevant threshold.
Thanks to the pre- and post-trigger recording,
we can maintain the sensitivity to the high
energy tail of the neutron spectrum.

Concerning the mechanical setup,
all system components will be loaded on a sturdy 2-tier
utility cart on wheels. The bottom shelf will host
the detector modules and the lead shield,
whereas the DAQ board, the front-end circuit,
the high voltage (HV) crate and the computer will
be placed on the top shelf.
The PMT and HV cables will be delivered from bottom
floor to top floor. The design of the complete mechanical
structure is ongoing.

%% file: summary.tex
\section{Summary}
\label{sec:Summary}

We presented the design of a mobile neutron spectrometer
for the LNGS underground laboratory. The detector consists
of 36 independent modules arranged in a 6-by-6 array. In each module, a plastic scintillator bar, after being wrapped
by reflector and gadolinium foils, is affixed
to a 3-inch PMT via optical glue.
The proton recoils
prompted by the thermalization
of a fast neutron in the scintillator
gives a measure of the neutron's
initial energy.
The high energy deposit following the
capture of the thermalized neutron on
gadolinium enables neutron tagging. The time
coincidence between the capture signal and
the preceding proton recoils lies
at the core of the design, which provides an excellent
background suppression. A further reduction
in background originating from accidental gammas
is accomplished by a \SI{16}{\mm} thick lead shield.
The detector construction is currently in progress.

We also studied the expected detector performance
within the Geant4 simulation framework. The simulations
indicate that the neutron threshold energy down to \SI{1}{\MeV}
can be feasibly reached. Using a simulated
$^{252}\mathrm{Cf}$ neutron source with a clipped
energy spectrum (1-10 MeV), the neutron detection
efficiency was found to be $\sim$6 \% at the capture
gamma energy threshold of \SI{3}{\MeV}.
We expect to detect about 12 neutrons per day,
assuming an ambient neutron flux of $0.42 \times 10^{-6}\,\mathrm{n/cm^2/s}$.

Two background sources were thoroughly investigated, namely
thermal neutrons
and the pile-up of environmental gammas. Thermal neutrons
will be captured on gadolinium sheets and
the capture pulses can accidentally pair up
with an ambient gamma, hence
faking fast neutron capture events. Based on the largest
thermal neutron flux number ever measured at LNGS,
$5.3 \times 10^{-6}\,\mathrm{n/cm^2/s}$,
we established that
the upper limit for
the thermal neutron-induced background rate
would be 2.26 counts per day.
Concerning the pile-up phenomenon,
two random environmental gammas can quasi-simultaneously
leave energy deposits in the detector
and the summed pulse
may be misidentified as a genuine capture pulse.
Combined with an accidental gamma within the
coincidence window, false neutron capture events may arise.
Assuming an ambient gamma flux of
1 $\gamma$$\mathrm{/cm^2/s}$, the pile-up-induced
background rate was conservatively estimated
to be about 0.45 events per day
at the capture signal threshold
of \SI{3}{\MeV}. The rate of ambient
gammas leaving at least \SI{50}{\keV} 
energy deposit (assumed threshold
for the preceding proton recoil signal)
in the detector is 218 Hz.
The probability of an ambient gamma
faking a proton recoil within the
coincidence time window is 2.16\%.

%% file: acknowledgements.tex
%***************************************************
% acknowledgements

\acknowledgments

We are grateful to
LNGS for their cooperation
and support throughout
the entire project
with special thanks to
Alba Formicola and
Axel Boeltzig.
We acknowledge
the financial support from the
German Federal Ministry of Education and Research (BMBF) under the grant number 05A21VK1
and the Italian National Institute
for Nuclear Physics (INFN).
We also thank Axel Boeltzig
for the comments on this manuscript.

%% file: main.bbl
\providecommand{\href}[2]{#2}\begingroup\raggedright\begin{thebibliography}{10}

\bibitem{WULANDARI2004313}
H.~Wulandari, J.~Jochum, W.~Rau and F.~{von Feilitzsch}, \emph{Neutron flux at
  the {G}ran {S}asso underground laboratory revisited},
  \href{http://dx.doi.org/https://doi.org/10.1016/j.astropartphys.2004.07.005}{\emph{Astroparticle
  Physics} {\bfseries 22} (2004) 313--322}.

\bibitem{Mei:2005gm}
D.~Mei and A.~Hime, \emph{{Muon-induced background study for underground
  laboratories}},
  \href{http://dx.doi.org/10.1103/PhysRevD.73.053004}{\emph{Phys. Rev. D}
  {\bfseries 73} (2006) 053004},
  [\href{https://arxiv.org/abs/astro-ph/0512125}{{\ttfamily
  astro-ph/0512125}}].

\bibitem{doi:10.1146/annurev.nucl.54.070103.181248}
J.~A. Formaggio and C.~Martoff, \emph{Backgrounds to sensitive experiments
  underground},
  \href{http://dx.doi.org/10.1146/annurev.nucl.54.070103.181248}{\emph{Annual
  Review of Nuclear and Particle Science} {\bfseries 54} (2004) 361--412}.

\bibitem{Bellotti:1985rz}
E.~Bellotti, M.~Buraschi, E.~Fiorini and C.~Liguori, \emph{New measurement of
  rock contaminations and neutron activity in the {G}ran {S}asso tunnel},
  Tech. Rep. INFN/TC-85/19, Italian National Institute for Nuclear Physics
  (INFN) Milan Unit, 1985.

\bibitem{Aleksan:1988qh}
R.~Aleksan et~al., \emph{Measurement of fast neutrons in the {G}ran {S}asso
  laboratory using a ${^6Li}$ doped liquid scintillator},
  \href{http://dx.doi.org/https://doi.org/10.1016/0168-9002(89)90380-X}{\emph{Nuclear
  Instruments and Methods in Physics Research Section A: Accelerators,
  Spectrometers, Detectors and Associated Equipment} {\bfseries 274} (1989)
  203--206}.

\bibitem{RINDI1988871}
A.~Rindi, F.~Celani, M.~Lindozzi and S.~Miozzi, \emph{Underground neutron flux
  measurement},
  \href{http://dx.doi.org/https://doi.org/10.1016/0168-9002(88)90772-3}{\emph{Nuclear
  Instruments and Methods in Physics Research Section A: Accelerators,
  Spectrometers, Detectors and Associated Equipment} {\bfseries 272} (1988)
  871--874}.

\bibitem{Belli1989}
P.~Belli et~al., \emph{Deep underground neutron flux measurement with large
  ${BF}_3$ counters}, \href{http://dx.doi.org/10.1007/BF02800162}{\emph{Il
  Nuovo Cimento A (1965-1970)} {\bfseries 101} (1989) 959--966}.

\bibitem{Cribier:1995zz}
M.~Cribier et~al., \emph{{Radiochemical measurement of fast neutrons using a
  ${Ca(NO_3)_2}$ aqueous solution}},
  \href{http://dx.doi.org/10.1016/0168-9002(95)00524-2}{\emph{Nucl. Instrum.
  Meth. A} {\bfseries 365} (1995) 533--541}.

\bibitem{Arneodo:1999py}
F.~Arneodo et~al., \emph{{Neutron background measurements in the Hall C of the
  Gran Sasso Laboratory}}, {\emph{Nuovo Cim. A} {\bfseries 112} (1999)
  819--831}.

\bibitem{DEBICKI2009429}
Z.~Dębicki et~al., \emph{Thermal neutrons at {G}ran {S}asso},
  \href{http://dx.doi.org/https://doi.org/10.1016/j.nuclphysbps.2009.09.084}{\emph{Nuclear
  Physics B - Proceedings Supplements} {\bfseries 196} (2009) 429--432}.

\bibitem{Haffke:2011fp}
M.~Haffke et~al., \emph{{Background measurements in the Gran Sasso Underground
  Laboratory}}, \href{http://dx.doi.org/10.1016/j.nima.2011.04.027}{\emph{Nucl.
  Instrum. Meth. A} {\bfseries 643} (2011) 36--41},
  [\href{https://arxiv.org/abs/1101.5298}{{\ttfamily 1101.5298}}].

\bibitem{BEST20161}
A.~Best et~al., \emph{Low energy neutron background in deep underground
  laboratories},
  \href{http://dx.doi.org/https://doi.org/10.1016/j.nima.2015.12.034}{\emph{Nuclear
  Instruments and Methods in Physics Research Section A: Accelerators,
  Spectrometers, Detectors and Associated Equipment} {\bfseries 812} (2016)
  1--6}.

\bibitem{Boeltzig2018}
A.~{Boeltzig} et~al., \emph{{Improved background suppression for radiative
  capture reactions at LUNA with HPGe and BGO detectors}},
  \href{http://dx.doi.org/10.1088/1361-6471/aaa163}{\emph{Journal of Physics G
  Nuclear Physics} {\bfseries 45} (Feb., 2018) 025203}.

\bibitem{DEBICKI2018133}
Z.~Dębicki et~al., \emph{Neutron flux measurements in the {G}ran {S}asso
  national laboratory and in the {S}lanic {P}rahova {S}alt {M}ine},
  \href{http://dx.doi.org/https://doi.org/10.1016/j.nima.2018.09.049}{\emph{Nuclear
  Instruments and Methods in Physics Research Section A: Accelerators,
  Spectrometers, Detectors and Associated Equipment} {\bfseries 910} (2018)
  133--138}.

\bibitem{Bruno2019}
G.~{Bruno} and W.~{Fulgione}, \emph{{Flux measurement of fast neutrons in the
  Gran Sasso underground laboratory}},
  \href{http://dx.doi.org/10.1140/epjc/s10052-019-7247-9}{\emph{European
  Physical Journal C} {\bfseries 79} (Sept., 2019) 747},
  [\href{https://arxiv.org/abs/1905.05512}{{\ttfamily 1905.05512}}].

\bibitem{Wulandari:2003px}
H.~Wulandari et~al., \emph{{Study on neutron induced background in the CRESST
  experiment}}, {\emph{IAU Symp.} {\bfseries 220} (2004) 491},
  [\href{https://arxiv.org/abs/hep-ex/0310042}{{\ttfamily hep-ex/0310042}}].

\bibitem{Abt:2004yk}
I.~Abt et~al., \emph{{A New $Ge^{76}$ Double Beta Decay Experiment at LNGS:
  Letter of Intent}},  \href{https://arxiv.org/abs/hep-ex/0404039}{{\ttfamily
  hep-ex/0404039}}.

\bibitem{BELLINI2010169}
F.~Bellini et~al., \emph{{M}onte {Ca}rlo evaluation of the external gamma,
  neutron and muon induced background sources in the {CUORE} experiment},
  \href{http://dx.doi.org/https://doi.org/10.1016/j.astropartphys.2010.01.004}{\emph{Astroparticle
  Physics} {\bfseries 33} (2010) 169--174}.

\bibitem{WRIGHT201118}
A.~Wright, P.~Mosteiro, B.~Loer and F.~Calaprice, \emph{A highly efficient
  neutron veto for dark matter experiments},
  \href{http://dx.doi.org/https://doi.org/10.1016/j.nima.2011.04.009}{\emph{Nuclear
  Instruments and Methods in Physics Research Section A: Accelerators,
  Spectrometers, Detectors and Associated Equipment} {\bfseries 644} (2011)
  18--26}.

\bibitem{XENON100:2013jxx}
{\scshape XENON100} collaboration, E.~Aprile et~al., \emph{{The neutron
  background of the XENON100 dark matter search experiment}},
  \href{http://dx.doi.org/10.1088/0954-3899/40/11/115201}{\emph{J. Phys. G}
  {\bfseries 40} (2013) 115201},
  [\href{https://arxiv.org/abs/1306.2303}{{\ttfamily 1306.2303}}].

\bibitem{XENON1T:2014eqx}
{\scshape XENON1T} collaboration, E.~Aprile et~al., \emph{{Conceptual design
  and simulation of a water Cherenkov muon veto for the XENON1T experiment}},
  \href{http://dx.doi.org/10.1088/1748-0221/9/11/P11006}{\emph{JINST}
  {\bfseries 9} (2014) P11006},
  [\href{https://arxiv.org/abs/1406.2374}{{\ttfamily 1406.2374}}].

\bibitem{CUORE:2017ztm}
{\scshape CUORE} collaboration, C.~Alduino et~al., \emph{{The projected
  background for the CUORE experiment}},
  \href{http://dx.doi.org/10.1140/epjc/s10052-017-5080-6}{\emph{Eur. Phys. J.
  C} {\bfseries 77} (2017) 543},
  [\href{https://arxiv.org/abs/1704.08970}{{\ttfamily 1704.08970}}].

\bibitem{Kim:2004wwa}
H.~J. Kim et~al., \emph{{Measurement of the neutron flux in the CPL underground
  laboratory and simulation studies of neutron shielding for WIMP searches}},
  \href{http://dx.doi.org/10.1016/j.astropartphys.2003.09.001}{\emph{Astropart.
  Phys.} {\bfseries 20} (2004) 549--557}.

\bibitem{Kudryavtsev:2008zzb}
{\scshape UKDMC, ZEPLIN-II, ILIAS} collaboration, V.~A. Kudryavtsev,
  \emph{{Neutron background in the Boulby Underground Laboratory}},
  \href{http://dx.doi.org/10.1088/1742-6596/120/4/042028}{\emph{J. Phys. Conf.
  Ser.} {\bfseries 120} (2008) 042028}.

\bibitem{2010arXiv1001.4383R}
S.~{Rozov} et~al., \emph{{Monitoring of the thermal neutron flux in the LSM
  underground laboratory}}, {\emph{arXiv e-prints} (Jan., 2010)
  arXiv:1001.4383}, [\href{https://arxiv.org/abs/1001.4383}{{\ttfamily
  1001.4383}}].

\bibitem{Eitel_2012}
K.~Eitel and the EDELWEISS~collaboration, \emph{Measurements of neutron fluxes
  in the {LSM} underground laboratory},
  \href{http://dx.doi.org/10.1088/1742-6596/375/1/012016}{\emph{Journal of
  Physics: Conference Series} {\bfseries 375} (jul, 2012) 012016}.

\bibitem{PARK2013302}
H.~Park, J.~Kim, Y.~Hwang and K.-O. Choi, \emph{Neutron spectrum at the
  underground laboratory for the ultra low background experiment},
  \href{http://dx.doi.org/https://doi.org/10.1016/j.apradiso.2013.03.068}{\emph{Applied
  Radiation and Isotopes} {\bfseries 81} (2013) 302--306}.

\bibitem{Jordan:2013exa}
D.~Jordan et~al., \emph{{Measurement of the neutron background at the Canfranc
  Underground Laboratory LSC}},
  \href{http://dx.doi.org/10.1016/j.astropartphys.2012.11.007}{\emph{Astropart.
  Phys.} {\bfseries 42} (2013) 1--6}.

\bibitem{ZENG2015108}
Z.~Zeng, H.~Gong, Q.~Yue and J.~Li, \emph{Thermal neutron background
  measurement in {CJPL}},
  \href{http://dx.doi.org/https://doi.org/10.1016/j.nima.2015.09.043}{\emph{Nuclear
  Instruments and Methods in Physics Research Section A: Accelerators,
  Spectrometers, Detectors and Associated Equipment} {\bfseries 804} (2015)
  108--112}.

\bibitem{Hu:2016vbu}
Q.~Hu et~al., \emph{{Neutron background measurements at China Jinping
  underground laboratory with a Bonner Multi-sphere Spectrometer}},
  \href{http://dx.doi.org/10.1016/j.nima.2017.03.048}{\emph{Nucl. Instrum.
  Meth. A} {\bfseries 859} (2017) 37--40},
  [\href{https://arxiv.org/abs/1612.04054}{{\ttfamily 1612.04054}}].

\bibitem{10.1093/ptep/pty133}
K.~Mizukoshi et~al., \emph{{Measurement of ambient neutrons in an underground
  laboratory at the Kamioka Observatory}},
  \href{http://dx.doi.org/10.1093/ptep/pty133}{\emph{Progress of Theoretical
  and Experimental Physics} {\bfseries 2018} (12, 2018) }.

\bibitem{Grieger2020}
M.~Grieger et~al., \emph{Neutron flux and spectrum in the {D}resden
  {F}elsenkeller underground facility studied by moderated $^{3}\mathrm{He}$
  counters}, \href{http://dx.doi.org/10.1103/PhysRevD.101.123027}{\emph{Phys.
  Rev. D} {\bfseries 101} (Jun, 2020) 123027}.

\bibitem{YOON2021102533}
Y.~S. Yoon, J.~Kim and H.~Park, \emph{Neutron background measurement for rare
  event search experiments in the {Y}ang{Y}ang underground laboratory},
  \href{http://dx.doi.org/https://doi.org/10.1016/j.astropartphys.2020.102533}{\emph{Astroparticle
  Physics} {\bfseries 126} (2021) 102533}.

\bibitem{Aglietta:1992dy}
M.~Aglietta et~al., \emph{{The Most powerful scintillator supernovae detector:
  LVD}}, \href{http://dx.doi.org/10.1007/BF02740929}{\emph{Nuovo Cim. A}
  {\bfseries 105} (1992) 1793--1804}.

\bibitem{BRAMBLETT1960395}
R.~L. Bramblett and T.~Bonner, \emph{Neutron evaporation spectra from (p, n)
  reactions},
  \href{http://dx.doi.org/https://doi.org/10.1016/0029-5582(60)90182-6}{\emph{Nuclear
  Physics} {\bfseries 20} (1960) 395--407}.

\bibitem{THOMAS200212}
D.~Thomas and A.~Alevra, \emph{Bonner sphere spectrometers—a critical
  review},
  \href{http://dx.doi.org/https://doi.org/10.1016/S0168-9002(01)01379-1}{\emph{Nuclear
  Instruments and Methods in Physics Research Section A: Accelerators,
  Spectrometers, Detectors and Associated Equipment} {\bfseries 476} (2002)
  12--20}.

\bibitem{FELDMAN1991350}
W.~Feldman, G.~Auchampaugh and R.~Byrd, \emph{A novel fast-neutron detector for
  space applications},
  \href{http://dx.doi.org/https://doi.org/10.1016/0168-9002(91)90342-N}{\emph{Nuclear
  Instruments and Methods in Physics Research Section A: Accelerators,
  Spectrometers, Detectors and Associated Equipment} {\bfseries 306} (1991)
  350--365}.

\bibitem{KAMYKOWSKI1992559}
E.~Kamykowski, \emph{Comparison of calculated and measured spectral response
  and intrinsic efficiency for a boron-loaded plastic neutron detector},
  \href{http://dx.doi.org/https://doi.org/10.1016/0168-9002(92)91002-Q}{\emph{Nuclear
  Instruments and Methods in Physics Research Section A: Accelerators,
  Spectrometers, Detectors and Associated Equipment} {\bfseries 317} (1992)
  559--566}.

\bibitem{AOYAMA1993492}
T.~Aoyama, K.~Honda, C.~Mori, K.~Kudo and N.~Takeda, \emph{Energy response of a
  full-energy-absorption neutron spectrometer using boron-loaded liquid
  scintillator {BC}-523},
  \href{http://dx.doi.org/https://doi.org/10.1016/0168-9002(93)91197-U}{\emph{Nuclear
  Instruments and Methods in Physics Research Section A: Accelerators,
  Spectrometers, Detectors and Associated Equipment} {\bfseries 333} (1993)
  492--501}.

\bibitem{ABDURASHITOV2002318}
J.~Abdurashitov et~al., \emph{A high resolution, low background fast neutron
  spectrometer},
  \href{http://dx.doi.org/https://doi.org/10.1016/S0168-9002(01)01447-4}{\emph{Nuclear
  Instruments and Methods in Physics Research Section A: Accelerators,
  Spectrometers, Detectors and Associated Equipment} {\bfseries 476} (2002)
  318--321}.

\bibitem{LANGFORD201578}
T.~Langford et~al., \emph{Fast neutron detection with a segmented
  spectrometer},
  \href{http://dx.doi.org/https://doi.org/10.1016/j.nima.2014.10.060}{\emph{Nuclear
  Instruments and Methods in Physics Research Section A: Accelerators,
  Spectrometers, Detectors and Associated Equipment} {\bfseries 771} (2015)
  78--87}.

\bibitem{Langford_2016}
T.~Langford, E.~Beise, H.~Breuer, C.~Heimbach, G.~Ji and J.~Nico,
  \emph{Development and characterization of a high sensitivity segmented fast
  neutron spectrometer ({F}a{NS}-2)},
  \href{http://dx.doi.org/10.1088/1748-0221/11/01/P01006}{\emph{Journal of
  Instrumentation} {\bfseries 11} (jan, 2016) P01006}.

\bibitem{ZAITSEVA2013747}
N.~Zaitseva et~al., \emph{Pulse shape discrimination with lithium-containing
  organic scintillators},
  \href{http://dx.doi.org/https://doi.org/10.1016/j.nima.2013.08.048}{\emph{Nuclear
  Instruments and Methods in Physics Research Section A: Accelerators,
  Spectrometers, Detectors and Associated Equipment} {\bfseries 729} (2013)
  747--754}.

\bibitem{BREUKERS201358}
R.~Breukers, C.~Bartle and A.~Edgar, \emph{Transparent lithium loaded plastic
  scintillators for thermal neutron detection},
  \href{http://dx.doi.org/https://doi.org/10.1016/j.nima.2012.10.080}{\emph{Nuclear
  Instruments and Methods in Physics Research Section A: Accelerators,
  Spectrometers, Detectors and Associated Equipment} {\bfseries 701} (2013)
  58--61}.

\bibitem{CHEREPY2015126}
N.~J. Cherepy et~al., \emph{Bismuth- and lithium-loaded plastic scintillators
  for gamma and neutron detection},
  \href{http://dx.doi.org/https://doi.org/10.1016/j.nima.2015.01.008}{\emph{Nuclear
  Instruments and Methods in Physics Research Section A: Accelerators,
  Spectrometers, Detectors and Associated Equipment} {\bfseries 778} (2015)
  126--132}.

\bibitem{Ellis2017}
E.~Mark~Ellis, C.~Hurlbut, C.~Allwork and B.~Morris, \emph{Neutron and gamma
  ray pulse shape discrimination with {EJ}-270 lithium-loaded plastic
  scintillator},  in \emph{2017 IEEE Nuclear Science Symposium and Medical
  Imaging Conference (NSS/MIC)}, pp.~1--5, 2017.
\newblock \href{http://dx.doi.org/10.1109/NSSMIC.2017.8532688}{DOI}.

\bibitem{Nemchenok2021}
I.~B. Nemchenok, I.~I. Kamnev, E.~A. Shevchik and I.~A. Suslov,
  \emph{Lithium-loaded plastic scintillators for the detection of thermal
  neutrons}, \href{http://dx.doi.org/10.3103/S1062873821050154}{\emph{Bulletin
  of the Russian Academy of Sciences: Physics} {\bfseries 85} (2021) 476--479}.

\bibitem{Potapov2015}
V.~N. Potapov et~al., \emph{A combined spectrometric detector of fast
  neutrons},
  \href{http://dx.doi.org/10.1134/S0020441215030094}{\emph{Instruments and
  Experimental Techniques} {\bfseries 58} (2015) 329--336}.

\bibitem{WILHELM2017}
K.~Wilhelm, J.~Nattress and I.~Jovanovic, \emph{Development and operation of a
  $^6{LiF}$:{Z}n{S}({A}g)—scintillating plastic capture-gated detector},
  \href{http://dx.doi.org/https://doi.org/10.1016/j.nima.2016.10.042}{\emph{Nuclear
  Instruments and Methods in Physics Research Section A: Accelerators,
  Spectrometers, Detectors and Associated Equipment} {\bfseries 842} (2017)
  54--61}.

\bibitem{BARTCZIRR1994532}
J.~{Bart Czirr} and G.~L. Jensen, \emph{A compact neutron coincidence
  spectrometer, its measured response functions and potential applications},
  \href{http://dx.doi.org/https://doi.org/10.1016/0168-9002(94)91222-X}{\emph{Nuclear
  Instruments and Methods in Physics Research Section A: Accelerators,
  Spectrometers, Detectors and Associated Equipment} {\bfseries 349} (1994)
  532--539}.

\bibitem{Nattress2016}
J.~Nattress, M.~Mayer, A.~Foster, A.~Barhoumi~Meddeb, C.~Trivelpiece,
  Z.~Ounaies et~al., \emph{Capture-gated spectroscopic measurements of
  monoenergetic neutrons with a composite scintillation detector},
  \href{http://dx.doi.org/10.1109/TNS.2016.2537761}{\emph{IEEE Transactions on
  Nuclear Science} {\bfseries 63} (2016) 1227--1235}.

\bibitem{HOLM201448}
P.~Holm, K.~Peräjärvi, S.~Ristkari, T.~Siiskonen and H.~Toivonen, \emph{A
  capture-gated neutron spectrometer for characterization of neutron sources
  and their shields},
  \href{http://dx.doi.org/https://doi.org/10.1016/j.nima.2014.03.021}{\emph{Nuclear
  Instruments and Methods in Physics Research Section A: Accelerators,
  Spectrometers, Detectors and Associated Equipment} {\bfseries 751} (2014)
  48--54}.

\bibitem{OVECHKINA2009}
L.~Ovechkina, K.~Riley, S.~Miller, Z.~Bell and V.~Nagarkar, \emph{Gadolinium
  loaded plastic scintillators for high efficiency neutron detection},
  \href{http://dx.doi.org/https://doi.org/10.1016/j.phpro.2009.07.008}{\emph{Physics
  Procedia} {\bfseries 2} (2009) 161--170}.

\bibitem{Dumazert2016}
J.~Dumazert, R.~Coulon, M.~Hamel, F.~Carrel, F.~Sguerra, S.~Normand et~al.,
  \emph{Gadolinium-loaded plastic scintillators for thermal neutron detection
  using compensation},
  \href{http://dx.doi.org/10.1109/TNS.2016.2535278}{\emph{IEEE Transactions on
  Nuclear Science} {\bfseries 63} (2016) 1551--1564}.

\bibitem{Poehlmann:2018sto}
D.~M. Poehlmann, D.~Barker, H.~Chagani, P.~Cushman, G.~Heuermann, A.~Medved
  et~al., \emph{{Characterization of Gadolinium-loaded Plastic Scintillator for
  Use as a Neutron Veto}},  \href{https://arxiv.org/abs/1812.11267}{{\ttfamily
  1812.11267}}.

\bibitem{PAWELCZAK2011}
I.~PaweŁczak, J.~Tõke, E.~Henry, M.~Quinlan, H.~Singh and W.~Schröder,
  \emph{Nstar—a capture gated plastic neutron detector},
  \href{http://dx.doi.org/https://doi.org/10.1016/j.nima.2010.11.103}{\emph{Nuclear
  Instruments and Methods in Physics Research Section A: Accelerators,
  Spectrometers, Detectors and Associated Equipment} {\bfseries 629} (2011)
  230--238}.

\bibitem{KURODA201241}
Y.~Kuroda et~al., \emph{A mobile antineutrino detector with plastic
  scintillators},
  \href{http://dx.doi.org/https://doi.org/10.1016/j.nima.2012.06.040}{\emph{Nuclear
  Instruments and Methods in Physics Research Section A: Accelerators,
  Spectrometers, Detectors and Associated Equipment} {\bfseries 690} (2012)
  41--47}.

\bibitem{MULMULE2018104}
D.~Mulmule et~al., \emph{A plastic scintillator array for reactor based
  anti-neutrino studies},
  \href{http://dx.doi.org/https://doi.org/10.1016/j.nima.2018.10.026}{\emph{Nuclear
  Instruments and Methods in Physics Research Section A: Accelerators,
  Spectrometers, Detectors and Associated Equipment} {\bfseries 911} (2018)
  104--114}.

\bibitem{Liu_2016}
Y.~Liu, Y.-G. Yang, Y.~Tai and Z.~Zhang, \emph{A capture-gated fast neutron
  detection method*},
  \href{http://dx.doi.org/10.1088/1674-1137/40/7/076201}{\emph{Chinese Physics
  C} {\bfseries 40} (jul, 2016) 076201}.

\bibitem{ET}
{ET Enterprises Limited}. \url{https://et-enterprises.com}.

\bibitem{Eljen}
{Eljen Technology}. \url{https://eljentechnology.com}.

\bibitem{3M}
{3M\texttrademark}. \url{https://www.3m.com}.

\bibitem{SAM}
{Stanford Advanced Materials}. \url{https://www.samaterials.com}.

\bibitem{GEANT4:2002zbu}
{\scshape GEANT4} collaboration, S.~Agostinelli et~al., \emph{{GEANT4--a
  simulation toolkit}},
  \href{http://dx.doi.org/10.1016/S0168-9002(03)01368-8}{\emph{Nucl. Instrum.
  Meth. A} {\bfseries 506} (2003) 250--303}.

\bibitem{EJ-500_abs}
J.~A. Green, A.~L. Guckes, J.~R. Tinsley and B.~J. Baldonado, \emph{{Gamma
  Array Simulation Toolkit for Modeling the NDSE Gamma Ray Detector Wall}},
  Tech. Rep. DOE/NV/03624-1339, Nevada National Security Site (NNSS), 2022.

\bibitem{Thulliez:2021ejj}
L.~Thulliez, C.~Jouanne and E.~Dumonteil, \emph{{Improvement of Geant4
  Neutron-HP package: From methodology to evaluated nuclear data library}},
  \href{http://dx.doi.org/10.1016/j.nima.2021.166187}{\emph{Nucl. Instrum.
  Meth. A} {\bfseries 1027} (2022) 166187},
  [\href{https://arxiv.org/abs/2109.05967}{{\ttfamily 2109.05967}}].

\bibitem{GLG4Sim}
{Generic liquid-scintillator anti-neutrino detector Geant4 simulation
  (GLG4sim)}, ``{Additional Gadolinium Support for GLG4sim}.''
  \url{https://www.phys.ksu.edu/personal/gahs/GLG4sim/Gd.html}, 2006.

\bibitem{LZ:2018qzl}
{\scshape LZ} collaboration, D.~S. Akerib et~al., \emph{{Projected WIMP
  sensitivity of the LUX-ZEPLIN dark matter experiment}},
  \href{http://dx.doi.org/10.1103/PhysRevD.101.052002}{\emph{Phys. Rev. D}
  {\bfseries 101} (2020) 052002},
  [\href{https://arxiv.org/abs/1802.06039}{{\ttfamily 1802.06039}}].

\bibitem{Solmaz:2020nhm}
M.~Solmaz, \emph{{Search for annual and diurnal modulations in the LUX
  experiment and assembling a tagged neutron source for the LZ Outer
  Detector}}.
\newblock PhD thesis, UC, Santa Barbara (main), 2020.

\bibitem{Weinmann2017}
R.~Weinmann-Smith, S.~Croft, M.~T. Swinhoe and A.~Enqvist, \emph{Changes to the
  $^{252}{C}f$ neutron spectrum caused by source encapsulation}, {\emph{ESARDA
  Bulletin} (2017) 44--53}.

\bibitem{Laplace:2020mfy}
T.~A. Laplace et~al., \emph{{Low Energy Light Yield of Fast Plastic
  Scintillators}},
  \href{http://dx.doi.org/10.1016/j.nima.2018.10.122}{\emph{Nucl. Instrum.
  Meth. A} {\bfseries 954} (2020) 161444},
  [\href{https://arxiv.org/abs/2009.07217}{{\ttfamily 2009.07217}}].

\bibitem{Birks1964}
J.~Birks, \emph{The Theory and Practice of Scintillation Counting}.
\newblock Pergamon, 1964.

\bibitem{Bellini:2009zw}
F.~Bellini et~al., \emph{{Monte Carlo evaluation of the external gamma, neutron
  and muon induced background sources in the CUORE experiment}},
  \href{http://dx.doi.org/10.1016/j.astropartphys.2010.01.004}{\emph{Astropart.
  Phys.} {\bfseries 33} (2010) 169--174},
  [\href{https://arxiv.org/abs/0912.0452}{{\ttfamily 0912.0452}}].

\bibitem{Bucci2008}
C.~Bucci et~al., \emph{Background study and {M}onte {C}arlo simulations for
  large-mass bolometers},
  \href{http://dx.doi.org/10.1140/epja/i2009-10805-7}{\emph{The European
  Physical Journal A} {\bfseries 41} (2009) 155--168}.

\bibitem{ARPESELLA1992420}
C.~Arpesella, \emph{Background measurements at {G}ran {S}asso {L}aboratory},
  \href{http://dx.doi.org/https://doi.org/10.1016/0920-5632(92)90207-9}{\emph{Nuclear
  Physics B - Proceedings Supplements} {\bfseries 28} (1992) 420--424}.

\bibitem{Malczewski2013}
D.~Malczewski, J.~Kisiel and J.~Dorda, \emph{Gamma background measurements in
  the {G}ran {S}asso {N}ational {L}aboratory},
  \href{http://dx.doi.org/10.1007/s10967-012-1990-9}{\emph{Journal of
  Radioanalytical and Nuclear Chemistry} {\bfseries 295} (2013) 749--754}.

\bibitem{Bemmerer:2005wu}
D.~Bemmerer et~al., \emph{{Feasibility of low energy radiative capture
  experiments at the LUNA underground accelerator facility}},
  \href{http://dx.doi.org/10.1140/epja/i2004-10135-4}{\emph{Eur. Phys. J. A}
  {\bfseries 24} (2005) 313--319},
  [\href{https://arxiv.org/abs/nucl-ex/0502007}{{\ttfamily nucl-ex/0502007}}].

\bibitem{Ahsan2021}
T.~Ahsan, C.~Swanson, T.~Qian, T.~Rubin and S.~Cohen, ``The pulse-pile-up tail
  artifact in pulse-height spectra.'' Unpublished manuscript available on
  webpage at \url{https://w3.pppl.gov/ppst/docs/ahsan.pdf}, July, 2021.

\bibitem{Tristan}
{TRISTAN group}, ``Conceptual design report: {KATRIN} with {TRISTAN} modules.''
  \url{https://www.katrin.kit.edu/downloads/TRISTAN__Technical_Design_Report\%20(10).pdf},
  2021.

\end{thebibliography}\endgroup
